\documentclass[fleqn,usenatbib]{rasti}

\usepackage{newtxtext,newtxmath}
 
\usepackage{graphicx}	% Including figure files

\usepackage[T1]{fontenc}
\usepackage{ae,aecompl}
\usepackage{lineno}
\usepackage{hyperref}
\usepackage{layout}
\usepackage[ruled,vlined]{algorithm2e}
\usepackage{svg}
\usepackage{academicons}
\usepackage{xcolor}

\usepackage{amsmath}	% Advanced maths commands
\usepackage{amssymb}	% Extra maths symbols

\definecolor{orange}{rgb}{1,0.5,0}
\definecolor{darkgreen}{rgb}{0,0.5,0}
\definecolor{codegreen}{rgb}{0,0.5,0}
\definecolor{codegray}{rgb}{0.5,0.5,0.5}
\definecolor{codepurple}{rgb}{0.58,0,0.82}
\definecolor{backcolour}{rgb}{1,1,1}
\usepackage{listings}

\lstdefinestyle{mystyle}{
    backgroundcolor=\color{backcolour},   
    commentstyle=\color{codegreen},
    keywordstyle=\color{codegreen},
    numberstyle=\tiny\color{codegray},
    stringstyle=\color{codepurple},
    basicstyle=\ttfamily,
    breakatwhitespace=false,         
    breaklines=true,                 
    captionpos=b,                    
    keepspaces=true,                 
    numbersep=10pt,                  
    showspaces=false,                
    showstringspaces=false,
    showtabs=true,                  
    tabsize=2,
    frame=tb
}
\newcommand{\Ngm}{N_{\mathrm{g}}}
\newcommand{\github}{\href{https://github.com/James11222/sarabande}{\textsc{sarabande github}}}

\newcommand{\orcid}[1]{
  \href{https://orcid.org/#1}{\includegraphics[width=10pt]{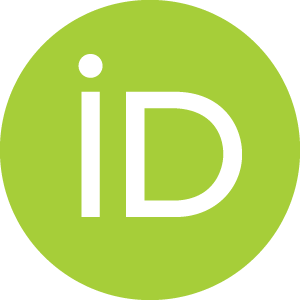}}
}

\title[\textsc{SARABANDE}: 3/4 PCFs with FFTs]{\textsc{SARABANDE}: 3/4 Point Correlation Functions with Fast Fourier Transforms}
\author[Sunseri et al.]{James Sunseri$^{1}$\orcid{0000-0003-4274-2662}\thanks{E-mail: jamessunseri@berkeley.edu (JS)},
Zachary Slepian$^{2}$\orcid{0000-0002-1208-119X}\thanks{E-mail: zslepian@ufl.edu (ZS)}, 
Stephen Portillo$^{3}$\orcid{0000-0001-8132-8056}\thanks{E-mail: sportill@uw.edu (SP)},
Jiamin Hou$^{2,4}$\orcid{0000-0001-6083-1947}\thanks{E-mail: jiamin.hou@ufl.edu (JH)},
\newauthor
{Sule Kahraman$^{5}$\orcid{0000-0002-3276-6030}\thanks{E-mail: sulekahraman97@gmail.com (SK)} \& Douglas P. Finkbeiner$^{6}$ \orcid{0000-0003-2808-275X}\thanks{E-mail: dfinkbeiner@cfa.harvard.edu (DPF)}}\\
\\
$^{1}$Department of Astronomy, University of California, Berkeley,
CA 94720-3411, USA\\
$^{2}$Department of Astronomy, University of Florida, 211 Bryant Space Science Center, Gainesville, FL 32611, USA\\
$^{3}$DIRAC Institute, Department of Astronomy, University of Washington, Seattle, WA 98195-1580, USA\\
$^{4}$Max-Planck-Institut f\"ur Extraterrestrische Physik, Postfach 1312, Giessenbachstrasse 1, 85748 Garching bei M\"unchen, Germany\\
$^{5}$Massachusetts Institute of Technology, Department of Electrical Engineering and Computer Science, 77 Massachusetts Avenue, Cambridge, MA 02139, USA\\
$^{6}$Center for Astrophysics | Harvard \& Smithsonian, 60 Garden Street, Cambridge, MA 02138, USA}
\date{Accepted XXX. Received YYY; in original form ZZZ} 

\pubyear{2022}

\begin{document}
\label{firstpage}
\pagerange{\pageref{firstpage}--\pageref{lastpage}}
\maketitle

% #########################################################
% #########################################################

%                        Abstract

% #########################################################
% #########################################################

\begin{abstract}
    We present a new \texttt{python} package \textsc{sarabande} for measuring 3 \& 4 Point Correlation Functions (3/4 PCFs) in $\mathcal{O} (\Ngm \log \Ngm)$ time using Fast Fourier Transforms (FFTs), with $\Ngm$ the number of grid points used for the FFT. \textsc{sarabande} can measure both projected and full 3 and 4 PCFs on gridded 2D and 3D datasets. The general technique is to generate suitable angular basis functions on an underlying grid, radially bin these to create kernels, and convolve these kernels with the original gridded data to obtain expansion coefficients about every point simultaneously. These coefficients are then combined to give us the 3/4 PCF as expanded in our basis. We apply \textsc{sarabande} to simulations of the Interstellar Medium (ISM) to show the results and scaling of calculating both the full and projected 3/4 PCFs. 
\end{abstract}

\begin{keywords}
methods: data analysis, statistical, numerical
\end{keywords}

% \section{To Do's}

% \begin{enumerate}
%     \item send to everyone for review
% \end{enumerate}

% #########################################################
% #########################################################

%                     Introduction

% #########################################################
% #########################################################

\section{Introduction}

Extracting structural and physical information from density and intensity maps is a common goal across multiple fields in astrophysics \citep{Peebles_Review}. Correlation functions are a natural means to characterize the translation and rotation-invariant information in a density field. The 2-Point Correlation Function (2PCF) or its Fourier-space analog the power spectrum are ubiquitous, and quantify correlations between pairs of points as a function of separation or Fourier-space wave-vector. However, for density fields beyond a Gaussian random field, they do not capture all of the information (\textit{e.g.} \citealt{2021MNRAS.505..628S}). 

To learn more about a given density field, one can look to higher-order statistics \citep{peebles2001galaxy}. The 3-Point Correlation Function (3PCF) and the 4-Point Correlation Function (4PCF) or their Fourier-space counterparts the bispectrum and trispectrum are the next statistics to consider \citep{peebles_fry_1978ApJ...221...19F, bitri_collis1998higher, MHD_burkhart_review_2010, MHD_TurbuStat,sabiu_2019ApJS..242...29S, trispectrum_2021}. These higher-order statistics measure correlations between triplets and quadruplets of points as a function of triangular and tetrahedral geometries in real or Fourier space respectively. They are useful for tracing non-Gaussian signals as well as probing parity-violation \citep{4PCF_odd_parity_cahn, 4PCF_odd_parity_jiamin}. The biggest hurdle for using these higher-order statistics has been the computational complexity. A naive calculation of an NPCF scales as $\mathcal{O}(N_{\mathrm{p}}^\mathrm{N})$ with $N_\mathrm{p}$ data points; this is infeasible for the ever-growing scale of astrophysical datasets. 

Recently a series of papers \citep{SE_3pt_alg, SE_3pt_FT, SE_aniso_3pt, philcox2021encore, NPCFDDimensions, hou2021analytic} have presented an approach to compute the 3PCF, 4PCF, 5PCF, and 6PCF that formally scales as $\mathcal{O}(N_{\mathrm{p}}^2)$ for $N_{\mathrm{p}}$ particles. It can be further accelerated using FFTs to $\mathcal{O}(N_{\rm g} \log N_{\rm g})$ with $N_{\rm g}$ the number of grid points used for the FFTs. In this work we present \textsc{sarabande}, a code package that computes the 3/4 PCFs using FFTs to obtain the coefficients needed in our estimator at all points simultaneously based on \citet{SE_3pt_alg} and \citet{philcox2021encore}.\footnote{Spherical-harmonic-based methods like those presented here have also been developed for the anisotropic 2-Point Correlation Function in \citet{SE_aniso_wa}, \citet{yama_ps} and for the power spectrum in \citet{Hand2017}; the base 3PCF algorithm of \citealt{SE_3pt_alg} is also implemented in \texttt{python} in \textsc{nbodykit} \citep{nbodykit}.} This method scales as $\mathcal{O}(N_{\rm g} \log N_{\rm g})$. \textsc{sarabande} is especially enabling for analysis of simulations that are already on regular grids, as FFTs require gridding but in this case, that will be lossless.  In principle \textsc{sarabande} could be used on any dataset as long as one is willing to grid it, and there are well-known methods to correct any artifacts introduced by gridding (\textit{e.g.} \citealt{jing}). We also present a version for 2D fields, suitable for computing the 3/4 PCF either on slices of gridded data or on a projected density field where one of the three axes has been integrated out.

This paper is structured as follows: Section \ref{sec:alg_in_brief} describes the underlying derivations and mathematical reasoning behind the FFT based approach to measuring the 3/4 PCFs. Section \ref{sec:code_struc} outlines how the code is implemented in \texttt{python} and how to use it. Section \ref{sec:results} discusses the overall performance of the algorithms and our process of validating the code. Section \ref{sec:Future Apps} offers potential future applications of \textsc{sarabande}. Section \ref{sec:conclusion} concludes. An Appendix describes the algorithms used in \textsc{sarabande}.

% \vspace{-0.1in}

% #########################################################
% #########################################################

%                   Algorithm in Brief

% #########################################################
% #########################################################

\section{The Algorithms}
\label{sec:alg_in_brief}
We here outline the derivations for the algorithms underlying \textsc{sarabande}. Some of the algorithms have been presented in previous works \citep{portillo2018, philcox2021encore, saydjari2021} but we present them here for completeness.

% #########################################################
%                       Full 3PCF
% #########################################################
\subsection{Full 3PCF}
\label{sec:3pcf} 

Here we briefly recapitulate the mathematical structure of the full 3PCF algorithm. More details of the base algorithm are in \cite{SE_3pt_alg}, and its implementation using FFTs in \cite{SE_3pt_FT}. 

The full 3PCF can be parameterized by two sides of a triangle $r_1$ and $r_2$ and the cosine of their enclosed angle, $\hat{r}_1 \cdot \hat{r}_2$. We can expand the full 3PCF, denoted $\zeta$, as a series of radial coefficients $\zeta_{\ell}$ dependent on $r_1$ and $r_2$ times the ``isotropic'' basis functions $\mathcal{P}_{\ell}(\hat{r}_1, \hat{r}_2)$ of \cite{Cahn2020IsotropicNB}, which are orthonormal and up to a phase and rescaling correspond to Legendre polynomials.\footnote{$\mathcal{P}_{\ell}(x) = (-1)^{\ell}\sqrt{2\ell + 1}/(4\pi) \mathcal{L}_{\ell}(x)$, with $\mathcal{L}_{\ell}$ the Legendre polynomial of order $\ell$.} We have
\begin{align}
\label{eq:eqn1}
    \zeta(r_1, r_2; \hat{r}_1 \cdot \hat{r}_2) = \sum_{\ell = 0}^{\infty} \zeta_{\ell}(r_1, r_2) \mathcal{P}_{\ell}(\hat{r}_1 \cdot \hat{r}_2).
\end{align}
We can form the full 3PCF as a volume-average ($V$ denotes volume) of ``local'' estimates of the full 3PCF (indicated by a hat) about points $\vec{x}$:
\begin{align}
    \zeta(r_1, r_2; \hat{r}_1 \cdot \hat{r}_2) = \int \frac{d^3 \vec{x}}{V}\; \hat{\zeta}(r_1, r_2; \hat{r}_1 \cdot \hat{r}_2; \vec{x}).
\end{align}
In turn, the local full 3PCF is constructed, for a density field $\delta$, as 
\begin{align}
\label{eqn:zeta_hat_z}
    \hat{\zeta}(r_1, r_2; \hat{r}_1 \cdot \hat{r}_2; \vec{x}) = \left< \delta(\vec{x}) \delta(\vec{x} + \vec{r}_1) \delta(\vec{x} + \vec{r}_2)\right>_{\mathcal{R}}
\end{align}
where the angle brackets with subscript $\mathcal{R}$ indicate an average over joint rotations $\mathcal{R}$ of $\hat{r}_1$ and $\hat{r}_2$ about $\vec{x}$. This rotation-averaging is lossless under the assumption of isotropy about $\vec{x}$. Since projection onto the isotropic basis functions is a linear operation, we then have that
\begin{align}
    \label{eqn:zeta_avg}
    \zeta_{\ell}(r_1, r_2) = \int \frac{d^3 \vec{x}}{V}\; \hat{\zeta}_{\ell}(r_1, r_2; \vec{x}).
\end{align}
The relation above means that the overall radial coefficients of the full 3PCF can be computed as the volume-average of the local estimates. We can write the local estimate of the multipole coefficients as
\begin{multline}
    \hat{\zeta}_{\ell}(r_1, r_2; \vec{x}) = \delta(\vec{x}) \int d \Omega_1 d\Omega_2 \; \mathcal{P}_{\ell}^*(\hat{r}_1, \hat{r}_2) \\ \times  \delta(\vec{x} + \vec{r_1}) \delta(\vec{x} + \vec{r_2}).
\end{multline}
Using equations 1 and 3 of \cite{Cahn2020IsotropicNB}, the basis functions for the full 3PCF can be written in terms of spherical harmonics:
\begin{align}
    \mathcal{P}_{\ell}(\hat{r}_1, \hat{r}_2) = \frac{ (-1)^{\ell}}{\sqrt{2 \ell + 1}} \sum_{m = -\ell}^{\ell} Y_{\ell m} (\hat{r}_1 )Y_{\ell m}^* (\hat{r}_2),
\end{align}
Combining these results, we can rewrite the local estimate of the multipole coefficients as
\begin{multline}
    \hat{\zeta}_{\ell}(r_1, r_2; \vec{x}) = \frac{(-1)^{\ell}}{\sqrt{2\ell + 1}} \delta(\vec{x}) \;  \\ \times \sum_{m = -\ell}^{\ell}  \int d\Omega_1 d\Omega_2 \; Y_{\ell m}^* (\hat{r}_1 )Y_{\ell m} (\hat{r}_2)  \delta(\vec{x} + \vec{r_1}) \delta(\vec{x} + \vec{r_2}).
\label{eqn:mu}
\end{multline}
The angular integrals may then be split; defining our coefficients as
\begin{align}
    a_{\ell m}(r_i; \vec{x}) = \int d\Omega_i\; Y_{\ell m}^*(\hat{r}_i) \delta(\vec{x} + \vec{r}_i),
\label{eqn:a_lm}
\end{align}
we find 
\begin{align}
    &\hat{\zeta}_{\ell}(r_1, r_2; \vec{x}) =  \frac{(-1)^{\ell}}{\sqrt{2\ell + 1}}\;\delta(\vec{x}) \sum_{m = - \ell}^{\ell} a_{\ell m}(r_1; \vec{x}) a_{\ell m}^*(r_2; \vec{x}).
\label{eqn:alm_to_zeta}
\end{align}
We notice that the key quantity is $a_{\ell m}$ as given by equation \ref{eqn:a_lm}, and that it has the structure of a convolution. Furthermore, since $a_{\ell m}$ has this structure, it may be computed around all points $\vec{x}$ at once using FFTs, leading to an algorithm scaling as $\mathcal{O} (N_{\rm g} \log N_{\rm g})$ with $N_{\rm g}$ as the number of grid points used for the FFT.

Generally we compute the full 3PCF radial coefficients on bins in $r_i$, and since binning is a linear operation, we can rewrite the $a_{\ell m}$ in equation \ref{eqn:a_lm} as 
\begin{align}
a_{\ell m}(r_i; \vec{x}) \to a^{\rm b}_{\ell m}( \vec{x}) = \int d\Omega \;Y_{\ell m}^*(\hat{r}) \delta^{\rm b}(\hat{r}; \vec{x})    
\label{eqn:discretize_almb}
\end{align}
where superscript $\rm b$ denotes a binned quantity, and we have defined the binned density field
\begin{align}
    \delta^{\rm b}(\hat{r}; \vec{x}) \equiv \int r^2 dr\; \delta(\vec{x} + \vec{r}) \Phi^{\rm b}(r). 
\label{eq:binned_delta}
\end{align}
$\Phi^{\rm b}$ is a binning function demanding that $r$ is in the $\rm b^{\rm th}$ bin.  We define the binning function to be a Heaviside function normalized by the volume of a given radial bin $V_{\mathrm{b}} = \int dr \; r^2 \Theta^{\rm b}(r)$ so that $\Phi^{\rm b}$ can be written as
\begin{align}
    \Phi^{\rm b}(r) = \frac{\Theta^{\rm b}(r)}{V_{\rm b}}.
\label{eq:bin_volume_unity}
\end{align}
With this binning scheme we can then write 
\begin{align}
\label{eqn:almconv}
    a^{\rm b}_{\ell m}(\vec{x}) = \int d^3\vec{r}\; Y_{\ell m}^*(\hat{r})\Phi^{\rm b}(r)\delta(\vec{x} + \vec{r}).
\end{align}
The above still clearly has the structure of a convolution and consequently can be obtained about all points $\vec{x}$ at once with an FFT.\footnote{equation \ref{eqn:almconv} technically has the form of a correlation due to the addition of $\vec{x}$ and $\vec{r}$ in $\delta$. In this work we also label it as a convolution to avoid confusion with the correlation coefficients.} Finally, we can rewrite equation \ref{eqn:alm_to_zeta} as
\begin{align}
\label{eqn:alm_to_zeta_clean}
    &\hat{\zeta}_{\ell}^{\; {\rm b}_1 {\rm b}_2}(\vec{x}) =  \frac{(-1)^{\ell}}{\sqrt{2\ell + 1}}\;\delta(\vec{x}) \sum_{m = - \ell}^{\ell} a_{\ell m}^{\rm b_1}(\vec{x}) a_{\ell m}^{\rm b_2 \; *}(\vec{x}),
\end{align}
where ${\rm b}_1$ and ${\rm b}_2$ represent the indices of the radial bins $r_1$ and $r_2$ fall into. The multipole coefficients $\zeta_{\ell}$ of the full 3PCF  are then constructed as the volume average of equation \ref{eqn:alm_to_zeta_clean}.

% #########################################################
%                     Projected 3PCF
% #########################################################

\subsection{Projected 3PCF}
\label{sec:proj 3pcf}

We now turn our attention to the projected 3PCF. Here the density is projected onto a 2D plane and therefore becomes a function of a 2D vector $\vec{x}$, which we then average over all 2D rotations. To do this we start with the full (projected) 3PCF
 As in Section \ref{sec:3pcf}, the triangle is parameterized by side lengths $\rho_i$ and the cosine of their enclosed angle: 
\begin{align}
\hat{\rho}_1 \cdot \hat{\rho}_2 \equiv \cos \phi_{12} = \cos (\phi_2 - \phi_1)
\end{align}
which we can rewrite as
\begin{align}
\cos (\phi_2 - \phi_1) = \frac{1}{2} \left[ e^{i(\phi_2 - \phi_1)} + e^{-i(\phi_2 - \phi_1)}\right].
\end{align}
Since the enclosed angle between the side lengths $\rho_1$ and $\rho_2$ can be rewritten in terms of complex exponentials, it is well motivated for us to expand the angle-dependence of the projected 3PCF in the basis of Fourier modes $\exp(im\phi_{12}$) yielding an analogous expression to equation \ref{eq:eqn1} in 2D:
\begin{align}
\zeta_{\rm proj}(\rho_1, \rho_2, \hat{\rho}_1 \cdot\hat \rho_2) = \sum_{m=0}^{\infty} \zeta_m(\rho_1, \rho_2) e^{im \phi_{12}}.
\end{align}
We can rewrite the basis functions as
\begin{align}
e^{im \phi_{12}} = e^{i m \phi_2} e^{-i m \phi_1}.
\end{align}
From here, the local projected 3PCF can be written as
\begin{align}
\label{eqn:zeta_hat}
    \hat{\zeta}_{\rm proj}(\rho_1, \rho_2; \hat{\rho}_1 \cdot \hat{\rho}_2; \vec{x}) = \left< \delta(\vec{x}) \delta(\vec{x} + \vec{\rho}_1) \delta(\vec{x} + \vec{\rho}_2)\right>_{\mathcal{R}},
\end{align}
where the angle brackets with subscript $\mathcal{R}$ indicate an average over joint 2D rotations $\mathcal{R}$ of $\hat{\rho}_1$ and $\hat{\rho}_2$ about $\vec{x}$ and $\delta$ is the projected density field. The projected 3PCF coefficients can be constructed by an area average of the "local" estimates of the projected 3PCF. In 2D we have
\begin{align}
 \zeta_{\rm proj}(\rho_1, \rho_2)  = \int \frac{d^2 \vec{x}}{A} \; \hat{\zeta}_{\rm proj}(\rho_1, \rho_2; \vec{x}),
\end{align}
and since projecting onto the Fourier basis is a linear operation, we have that 
\begin{align}
 \zeta_m(\rho_1, \rho_2)  = \int \frac{d^2 \vec{x}}{A} \; \hat{\zeta}_m(\rho_1, \rho_2; \vec{x}).    
\end{align}
The radial coefficients of the projected 3PCF can be computed as an area average of the local estimates. The local estimate of the multipole coefficients can be written as
\begin{multline}
\hat{\zeta}_m(\rho_1, \rho_2; \vec{x}) = \int d\phi_1 d\phi_2\;  e^{-i m \phi_2} e^{i m \phi_1} \\ \times \delta(\vec{x}) \delta(\vec{x} + \vec{\rho}_1) \delta(\vec{x} + \vec{\rho}_2). 
\end{multline}
Using orthogonality of the Fourier modes to project onto $\exp(i m\phi)$, we must integrate against its conjugate. We define 
\begin{align}
\label{eq:c_m def}
c_m(\rho_i; \vec{x}) \equiv \int d\phi\; e^{-i m \phi} \delta(\vec{x} + \vec{\rho}_i),
\end{align} 
which has the structure of a convolution. We exploit the fact that $c_{-m} = c_m^*$ as can be seen from making this replacement in equation \ref{eq:c_m def}. These convolution coefficients can be directly measured with \textsc{sarabande}. The $c_m(\rho_i;\vec{x})$ are the projected equivalent of the $a_{\ell m}(r_i, \vec{x})$. From this, we can write $\hat{\zeta}_m$ as their product:  
\begin{equation}
    \hat{\zeta}_{m}(\rho_1, \rho_2;\vec{x}) =  \frac{\delta(\vec{x})}{(2\pi)^2} \; c_m(\rho_1; \vec{x}) c_m^*(\rho_2; \vec{x}).
\end{equation}
Now we may proceed analogously to our work in equation \ref{eqn:discretize_almb} to bin these coefficients.
\begin{equation}
\label{eq:discretize_cmb}
     c_{m}(\rho_i;\vec{x}) \to c_{m}^{\rm b}(\vec{x}) = \int \delta^{\rm b}(\hat{\rho}_i ; \vec{x}) e^{-im \phi} d\phi,
\end{equation}
where the binned density field is
\begin{equation}
     \delta^{\rm b}(\hat{\rho}_i ;\vec{x}) \equiv \int \rho_i  d\rho_i \; \delta(\vec{x} + \vec{\rho}_i) \Phi^{\rm b}(\rho_i),
\label{eq:binned_delta_2D}
 \end{equation}
 and $\Phi^{\rm b}$ is a binning function. We define this binning function to be a Heaviside function normalized by the area of a given radial bin $A_{\mathrm{b}} = \int dr \; \Theta^{\rm b}(r)$ so that $\Phi^{\rm b}$ can be written as
\begin{align}
    \Phi^{\rm b}(r) = \frac{\Theta^{\rm b}(r)}{A_{\rm b}}.
\label{eq:bin_area_unity}
\end{align}
This binning function demands that $\rho_i$ is in a normalized radial bin having integral unity. With the binning function we can write the multipole coefficients of the projected 3PCF as 
\begin{align}
\hat{\zeta}_{m}^{\mathrm{b_1} \mathrm{b_2}}(\vec{x}) = \frac{\delta(\vec{x})}{(2\pi)^2} \; c_m^{\rm b_1}(\vec{x}) c_m^{\rm b_2 \; *}(\vec{x}),
\end{align}
where $\rm b_1$ and $\rm b_2$ represent the indices of the radial bins $\rho_1$ and $\rho_2$ fall into. If we sum over all values of $m$ we construct the projected 3PCF
\begin{align}
\zeta_{\rm proj}^{\mathrm{b_1} \mathrm{b_2}}(\hat{\rho}_1 \cdot\hat \rho_2; \vec{x}) = \frac{\delta(\vec{x})}{(2\pi)^2} \sum_{m=0}^{\infty} \; c_m^{\rm b_1}(\vec{x}) c_m^{\rm b_2 \; *}(\vec{x}) e^{im \phi_{12}}.
\label{eq: proj_3pcf_coeff}
\end{align}
These $c_m^{\rm b}$ coefficients can be interpreted as the convolution of the density field with the kernel 
\begin{equation}
    K_{m}^{\rm b}(\vec{\rho}_i) \equiv \Phi^{\rm b}(\rho_i) e^{-im \phi}.
\end{equation}
We can thus write the $c_m^{\rm b}$ coefficients as 
\begin{equation}
\label{eqn: c_m def kernel}
    c_{m}^{\rm b}(\vec{x}) = \left[\delta(\vec{\rho}_i) \star K_{m}^{\rm b}(\vec{\rho}_i) \right](\vec{x}).
\end{equation}
We define the $\star$ symbol to denote a convolution.

% #########################################################
%                         Full 4PCF
% #########################################################

\subsection{Full 4PCF}

% [here we briefly discuss the 4PCF derivation, for further details see philcox and isotropic basis stuff]
Here we briefly discuss the derivation of the 4PCF. Further details are given on the derivation in \cite{philcox2021encore} and further information on the isotropic basis functions can be found in \cite{Cahn2020IsotropicNB}. The formalism of the full 4PCF is similar to that of the full 3PCF; now we just have four points in the estimator as opposed to three. This translates to a tetrahedral geometry instead of a triangular one. Instead of just two vectors: we now have three $\vec{r}_1$,  $\vec{r}_2$, and $\vec{r}_3$. We can write the full 4PCF in its most basic form as
\begin{equation}
    \label{eqn:full 4pcf}
    \zeta = \langle \delta(\vec{x}) \delta(\vec{x} + \vec{r}_1) \delta(\vec{x} + \vec{r}_2) \delta(\vec{x} + \vec{r}_3) \rangle_{\mathcal{R}}.
\end{equation}
We have left unstated the arguments on the left-hand side intentionally as there are different ways to parameterize the full 4PCF, as we will shortly discuss. We can characterize the tetrahedron by the lengths $r_1$, $r_2$, and $r_3$ and the angles between $\vec{r}_1$,  $\vec{r}_2$, and $\vec{r}_3$. The dependence on these angles can be decomposed in the basis of isotropic functions developed by  \citet{Cahn2020IsotropicNB}. We then have the full 4PCF as a sum of radial coefficients $\zeta_{\Lambda}$ times these isotropic basis functions $\mathcal{P}_{\Lambda}$ for the angular dependence:
\begin{equation}
\label{4pcf_basis}
    \zeta = \sum_{\Lambda} \hat{\zeta}_{\Lambda}^{R} \mathcal{P}_{\Lambda}(\hat{R}).
\end{equation}
We define $\Lambda \equiv \{\ell_1, \ell_2, \ell_3\}$ to be the set of angular momenta describing the angles of the quadruplets used for the full 4PCF. The isotropic orthonormal basis used in equation \ref{4pcf_basis} is defined as
\begin{equation}
    \mathcal{P}_{\Lambda}(\hat{R}) = \sum_{M} C^{\Lambda}_{M} Y_{\ell_1 m_1}(\hat{r}_1) Y_{\ell_2 m_2}(\hat{r}_2) Y_{\ell_3 m_3}(\hat{r}_3),
\end{equation}
and $C^{\Lambda}_{M}$ is given by the Wigner 3-$j$ symbol with an additional phase:
\begin{equation}
    C^{\Lambda}_{M} = (-1)^{\ell_1 + \ell_2 + \ell_3} \begin{pmatrix}
\ell_1 & \ell_2 & \ell_3\\
m_1 & m_2 & m_3
\end{pmatrix}.
\end{equation}
For compactness we have defined $M \equiv \{m_1, m_2. m_3\}$, $\hat{R} \equiv \{ \hat{r}_1, \hat{r}_2, \hat{r}_3\}$, and $R \equiv \{ r_1, r_2, r_3 \}$. 
Equation \ref{4pcf_basis} shows the radial coefficients $\hat{\zeta}_{\Lambda}^{R}$ must be computed to compute the full 4PCF. This may be done in terms of binned convolution coefficients $a_{\ell m}^{\rm b}(\vec{x})$ defined in equation \ref{eqn:almconv}, leading to the radial coefficients about each point $\vec{x}$ as
% \begin{equation}
%     \hat{\zeta}_{\ell_1 \ell_2 \ell_3}^{\rm b_1 \rm b_2 \rm b_3}(\vec{x}) = \delta(\vec{x})\sum_{M} \mathcal{E}(\Lambda) \; C^{\Lambda}_{M} \; a_{\ell_1 m_1}^{\rm b_1}(\vec{x}) \; a_{\ell_2 m_2}^{\rm b_2}(\vec{x}) \; a_{\ell_3 m_3}^{\rm b_3}(\vec{x}),
% \end{equation}
\begin{multline}
    \hat{\zeta}_{\ell_1 \ell_2 \ell_3}^{\; \rm b_1 \rm b_2 \rm b_3}(\vec{x}) = \delta(\vec{x})\sum_{M} \mathcal{E}(\Lambda) \; C^{\Lambda}_{M} \\ \times \; a_{\ell_1 m_1}^{\rm b_1}(\vec{x}) \; a_{\ell_2 m_2}^{\rm b_2}(\vec{x}) \; a_{\ell_3 m_3}^{\rm b_3}(\vec{x}),
\label{eqn:mu}
\end{multline}
where $\mathcal{E}(\Lambda)$ is +1 for even parity and -1 for odd parity. Even parity occurs when $\ell_1 + \ell_2 + \ell_3$ is an even number, and odd parity occurs when the sum of the angular momenta is odd. For further details on this derivation, see Section 3.2 of \cite{philcox2021encore}.

% #########################################################
%                     Projected 4PCF
% #########################################################

\subsection{Projected 4PCF}
\label{sec:proj_4pcf}

For the projected 4PCF, we take a slightly different approach from that of the previous subsections. We begin by setting up a coordinate system whose origin is defined by one particle, which is always allowed by translational invariance. We then expand the dependence on $\vec{\rho}_i$ for the three remaining points in polar coordinates: 
\begin{gather}
\label{eq:proj 4pcf}
\delta(\vec{x}) \delta(\vec{x} + \vec{\rho}_1)  \delta(\vec{x} + \vec{\rho}_2)  \delta(\vec{x} + \vec{\rho}_3) = \\ \nonumber \delta(\vec{x}) \sum_{M} c_{m_1}(\rho_1;\vec{x}) e^{i m_1 \phi_1}
c_{m_2}(\rho_2;\vec{x}) e^{i m_2 \phi_2}
c_{m_3}(\rho_3;\vec{x}) e^{i m_3 \phi_3}.
\end{gather}
This can be done because the Fourier modes are a basis for any function of $\phi$. We capture the radial dependence on $\rho_i$ with the coefficients $c_{m_i}$. These coefficients were previously defined in Section \ref{sec:proj 3pcf} (see equation \ref{eq:c_m def}). We let $M \equiv \{m_1, m_2, m_3 \}$ for compactness as done in Section \ref{sec:proj 3pcf}. 

We now consider the constraints required for the projected 4PCF to be rotation-invariant. The "local" estimates of the projected 4PCF are found by averaging over joint rotations as done in the previous sections:
\begin{multline}
    \hat{\zeta}_{\mathrm{proj}}(\rho_1, \rho_2, \rho_3; \hat{\rho}_1, \hat{\rho}_2, \hat{\rho}_3;  \vec{x}) = \\ \langle \delta(\vec{x}) \delta(\vec{x} + \vec{\rho}_1)  \delta(\vec{x} + \vec{\rho}_2)  \delta(\vec{x} + \vec{\rho}_3) \rangle_{\mathcal{R}}.
\end{multline}
Returning to equation \ref{eq:proj 4pcf}, we can perform the average over rotations by adding an arbitrary displacement in angle, $\phi$, to each $\phi_i$ and integrating over it. We find
\begin{multline}
\hat{\zeta}_{\mathrm{proj}}(\rho_1, \rho_2, \rho_3; \hat{\rho}_1, \hat{\rho}_2, \hat{\rho}_3;  \vec{x}) = \\  \qquad \delta(\vec{x}) \sum_{M} c_{m_1}(\rho_1;\vec{x}) c_{m_2}(\rho_2;\vec{x}) c_{m_3}(\rho_3;\vec{x}) \; \\  \times \frac{1}{2\pi} \int d\phi\;  e^{i m_1( \phi_1 + \phi)}
 e^{i m_2 (\phi_2 + \phi)}
 e^{i m_3(\phi_3+ \phi)}. 
\end{multline}
From here we may rewrite the angular integral by factoring the exponentials
\begin{multline}
\frac{1}{2\pi} \int d\phi\;  e^{i m_1( \phi_1 + \phi)}
 e^{i m_2 (\phi_2 + \phi)}
 e^{i m_3(\phi_3+ \phi)} = \\ 
  \left( e^{i m_1 \phi_1} e^{i m_2 \phi_2} e^{i m_3 \phi_3} \right) \; \frac{1}{2 \pi} \int d\phi\; e^{i(m_1 + m_2 +m_3) \phi}. 
\end{multline}
The integral is only non-zero if $m_1 + m_2 + m_3 = 0$. Thus we may replace $m_{3} = -(m_1 + m_2)$. We then have the local estimate of the coefficient of the rotation-averaged projected 4PCF as 
\begin{multline}
\hat{\zeta}_{\mathrm{proj}} (\rho_1, \rho_2, \rho_3; \vec{x} ) = \\ \qquad \delta(\vec{x}) \sum_{m_1,\; m_2} c_{m_1}(\rho_1;\vec{x}) c_{m_2}(\rho_2;\vec{x}) c_{-(m_1+m_2)}(\rho_3;\vec{x})\\ \times \left( e^{i m_1 \phi_1} e^{i m_2 \phi_2} 
 e^{-i(m_1+m_2) \phi_3} \right).
\end{multline}
From here we reuse equation \ref{eq:discretize_cmb} to bin the convolution coefficients $c_{m_i}$. For compactness we define $\phi_{13} \equiv \phi_1 - \phi_3$ and $\phi_{23} \equiv \phi_{2} - \phi_{3}$. Finally, our local coefficient estimate of the projected 4PCF becomes
\begin{gather}
\hat{\zeta}_{\mathrm{proj}}^{\mathrm{b_1} \mathrm{b_2} \mathrm{b_3}}(\hat{\rho}_1,\hat{\rho}_2,\hat{\rho}_3; \vec{x} ) = \\ \nonumber \qquad \delta(\vec{x}) \sum_{m_1,\; m_2} c_{m_1}^{\rm b_1}(\vec{x}) c_{m_2}^{\rm b_2}(\vec{x}) c_{-(m_1+m_2)}^{\rm b_3}(\vec{x}) \left( e^{i m_1 \phi_{13}} e^{i m_2 \phi_{23}}  \right).
\end{gather}
The projected 4PCF coefficients are then
\begin{equation}
\hat{\zeta}_{m_1 m_2}^{\; \mathrm{b_1} \mathrm{b_2} \mathrm{b_3}}(\vec{x} ) = \delta(\vec{x}) c_{m_1}^{\rm b_1}(\vec{x}) c_{m_2}^{\rm b_2}(\vec{x}) c_{-(m_1+m_2)}^{\rm b_3}(\vec{x}).
\label{eq: proj_4pcf_coeff}
\end{equation}

\subsection{Normalization}
\label{sec:normalization}
Here we discuss the normalization of the FFT-based NPCF. We need to normalize because by definition the NPCF is a measure of the excess probability over random of finding a particle N-tuplet  (\textit{e.g.} the 2PCF is the probability excess of finding a galaxy pair over a random distribution). In order to normalize the measured NPCF coefficients to have this meaning we must follow an analogous normalization scheme to that of codes such as \textsc{encore}.\footnote{\textsc{encore} \citep{philcox2021encore} uses a generalization of the Landy-Szalay estimator wherein random particles enable subtracting the mean and dividing by it. This is instead of working with the density contrast $\delta$ directly. The particle-based normalization of \textsc{encore} includes the number density to represent the number of neighbors in a bin around a primary galaxy: $\mathcal{N} = N_{\rm gal} \prod_{i=1}^{N-1}(\bar{n}  V_{\rm bin,i})$, with $V_{\rm bin}=(4\pi/3) (r_+^3-r_-^3)$ with $r_+$ and $r_-$ respectively the upper and lower bounds of a given spherical shell.} Typical particle-based NPCF codes measure the correlations on a set of $N_{\rm gal}$ galaxies within a cosmological volume using an estimator. Our grid-based code does not have a distribution of particles; instead, \textsc{sarabande} takes a density contrast field $\delta$ as its input. We define $\delta = n/\bar{n} - 1$ where $n$ is the number density and $\bar{n}$ is the average number density.

\subsubsection{Normalizing the Spherical-Shell Bins}
In \textsc{sarabande}, the full 3/4 PCFs are measured on 3D bins in the side lengths of the configuration about the primary. Each bin is normalized to have volume unity as stated in equation \ref{eq:bin_volume_unity}. This normalization must be adjusted to reflect the fact that the underlying space is on a grid. Here we show how to perform this normalization. Since \textsc{sarabande} is computed using a grid mesh, we count the number of cells within a radial bin and multiply that by the volume of each cell. The volume of each cell in 3D is $V_{\rm cell} = (L/N_{\rm mesh})^3$ where $L$ is the physical box size of our data. Therefore the bin volume is defined as 
\begin{equation}
    V_{\mathrm{bin}} = V_{\rm cell} \times N_{\textrm{cells in bin}} = \left(\frac{L}{N_{\rm mesh}}\right)^3 N_{\textrm{cells in bin}}.
    \label{eqn:radial_binning}
\end{equation}
The text below equation \ref{eq:binned_delta} has additional information regarding our binning scheme. We will divide the full 3/4 PCF coefficients by the product of bin volumes to normalize as discussed below.

\subsubsection{Total Number of Objects}
\label{subsubsec: TNO}
Particle-based codes such as \textsc{encore} typically normalize by the total number of galaxies $N_{\rm gal}$ in a cosmological volume.
For particle-based codes this is simply the number of objects (or their weighted sum, if weights are used such as might be done to correct for survey systematics or survey geometry), but for density fields the number of particles used is not as straightforward. If we assume the input density field to have a mean of zero, then the equivalent value is
\begin{equation}
    N_{\rm gal} = \sum_i\sum_j\sum_k [\delta(\vec{x}) + 1].
    \label{eq:num_gals}
\end{equation}
In essence, we sum the density field over all the cells in the mesh, which are indexed by ($i,j,k$), to get the equivalent of the number of particles.

To properly normalize the full 3/4 PCF coefficients in equation \ref{eqn:alm_to_zeta_clean} and equation \ref{eqn:full 4pcf} we divide the coefficients by a normalization coefficient $\mathcal{N}$ which normalizes our bins to have volume unity and allows us to compare our coefficients to particle-based codes such as \textsc{encore} as done in Figure \ref{fig:full_compare}. The final normalization coefficient $\mathcal{N}$ reads
\begin{equation}
    \mathcal{N} = N_{\rm gal} \prod_{n=1}^{N-1} \bar{n} \; V_{\mathrm{bin},n}.
\end{equation}
We define $N$ is the order of the correlation function (\textit{e.g.} $N = 3$ for the 3PCF). We include the average number density $\bar{n} \equiv N_{\rm gal} / L^3$ in the product to represent the number of neighbors in a bin around primary point.

\subsubsection{Projected Case}
If we follow the same normalization procedure as the full 3/4 PCF for the projected 3/4 PCF, we find that the main difference is that instead of volumes in 3D we are working with areas in 2D. In \textsc{sarabande}, the projected 3/4 PCFs are measured on 2D bins in the side lengths of the configuration about the primary. Each bin is normalized to have area unity as stated in equation \ref{eq:bin_area_unity}. This normalization must be adjusted to reflect the fact that the underlying space is on a grid. Here we show how to perform this normalization. We still have a grid mesh in the projected case so to get the total area of a bin we have multiply the area of a cell by the number of cells within a bin. The area of a cell in 2D is $A_{\rm cell} = (L/N_{\rm mesh})^2$. Therefore the bin area is defined as
\begin{equation}
    A_{\mathrm{bin}} = A_{\rm cell} \times N_{\textrm{cells in bin}} = \left(\frac{L}{N_{\rm mesh}}\right)^2 N_{\textrm{cells in bin}}.
    \label{eq: proj_b_area}
\end{equation}
The text below equation \ref{eq:binned_delta_2D} has additional information regarding our binning scheme in 2D. 

We follow the same procedure for creating an equivalent expression for the number of particles as in equation \ref{eq:num_gals}. In 2D we find
\begin{equation}
    N_{\rm gal} = \sum_i \sum_j [\delta(\vec{x}) + 1],
\end{equation}
where we sum the density field over all the cells in the mesh indexed by ($i,j$). If we assume the density field has a mean of zero. 

\begin{figure*}
\centering
\includegraphics[width=2\columnwidth]{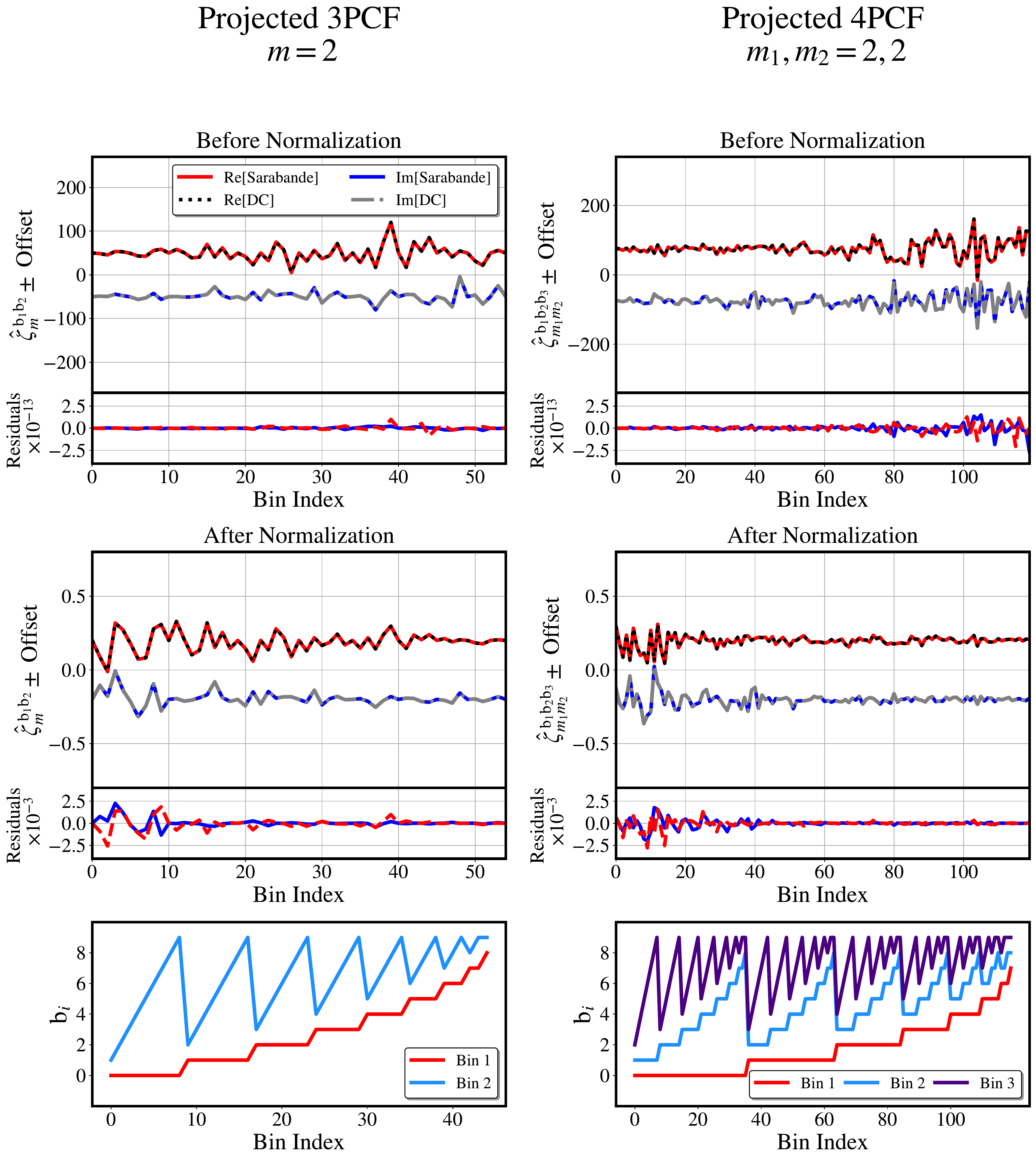}
\caption{This figure shows the agreement between \textsc{sarabande} and a particle-based direct counting (DC) code before and after normalization discussed in Section \ref{sec:normalization} (equations \ref{eq: proj_b_area}-\ref{eq: proj_norm}). The left column displays the projected 3PCF with $m = 2$ as an example of how normalization impacts the agreement between \textsc{sarabande} and particle-based codes. The right column shows the projected 4PCF with $m_1 = 2$ and $m_2 = 2$ as an example. For all plots of projected 3/4 PCF coefficients $\hat{\zeta}$ we add an arbitrary offset to the real parts and subtract the offset from the imaginary parts for clarity. For all residual plots the red dashed curve shows the residual of the real part of the projected 3/4 PCF coefficients and the blue solid curve shows the residual of the imaginary part. Before normalization \textsc{sarabande} has an agreement precision of $\sim 10^{-13}$ with the direct-counting code, after normalization \textsc{sarabande} has an agreement precision of $\sim 10^{-3}$ for this choice of resolution. This precision after normalization improves as we increase the resolution. The bottom row shows how we map the 2 or 3 radial bins to a 1D bin index. For 2 radial bins we have a nested \texttt{for} loop where $\textrm{b}_1$ iterates from 0 to $N_{\rm bins}$ then $\textrm{b}_2$ iterates from $\textrm{b}_1 + 1$ to $N_{\rm bins}$ (we avoid bin overlap where $\textrm{b}_1 = \textrm{b}_2$). Each iteration step is assigned to the combination ($\textrm{b}_1$, $\textrm{b}_2$). For 3 radial bins we add an additional nested \texttt{for} loop for $\textrm{b}_3$. With 3 radial bins we iterate through the bin combinations such that $\textrm{b}_1$ iterates from 0 to $N_{\rm bins}$ and $ \textrm{b}_1 + 1 < \textrm{b}_2 + 1 < \textrm{b}_3 < N_{\rm bins}$ to avoid bin overlap. Each combination of ($\textrm{b}_1$, $\textrm{b}_2$, $\textrm{b}_3$) is assigned to an iteration index. We iterate with these conditions so that we only display the unique correlation coefficients. We plot the coefficients as a function of 1D bin index to show all the 3/4 PCF coefficients measured. The input data used for this figure was a set of 41 randomly placed particles on a grid with a resolution of $N_{\rm g} = 256^2$ grid cells. To create the particles on the grid we assign cells within the grid a weight equal to unity. We use a small sample of particles because of the slow nature of the direct counting method. For this measurement we use 10 linearly-spaced radial bins.}
\label{fig:normalization_projected}
\end{figure*}

To properly normalize the projected 3/4 PCF coefficients in equation \ref{eq: proj_3pcf_coeff} and equation \ref{eq: proj_4pcf_coeff} we divide the coefficients by a normalization coefficient $\mathcal{N_{\rm proj}}$ which normalizes our bins to have area unity. This allows us to compare our coefficients to particle-based codes as done in Figure \ref{fig:normalization_projected}. In Section \ref{sec:code_use}, we describe the code including an optional flag to turn on this normalization procedure. The final normalization coefficient $\mathcal{N_{\rm proj}}$ reads
\begin{equation}
    \mathcal{N_{\rm proj}} = N_{\rm gal} \prod_{n=1}^{N-1} \bar{n} \; A_{\mathrm{bin},n}.
    \label{eq: proj_norm}
\end{equation}
We define $N$ to be the order of the correlation function as done at the end of Section \ref{subsubsec: TNO}. We include the average number density $\bar{n} \equiv N_{\rm gal} / L^2$ in the product to represent the number of neighbors in a bin around primary point.

\begin{figure}
\includegraphics[width=\columnwidth]{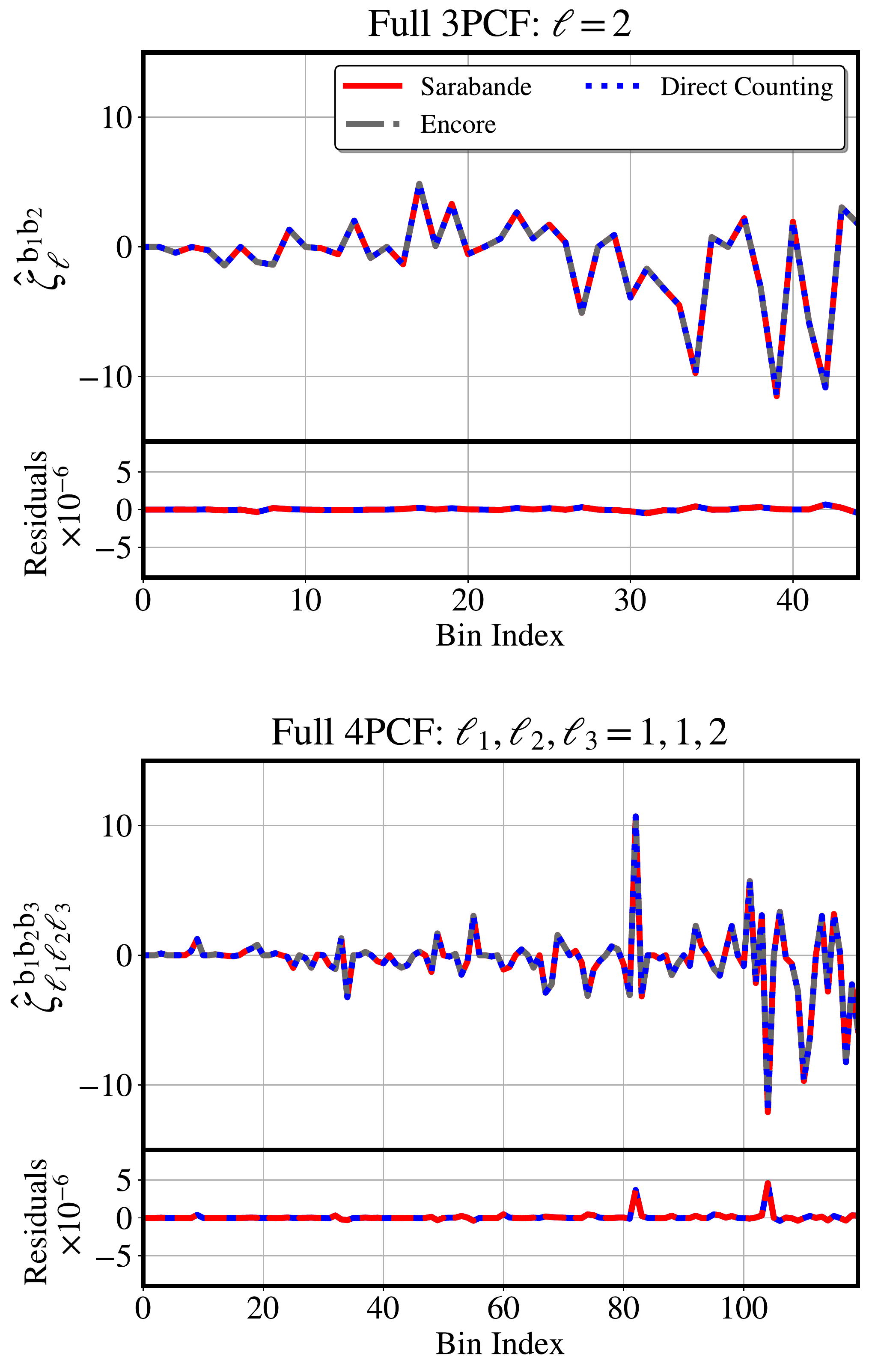}
\centering
\caption{This figure shows the accuracy of \textsc{sarabande} relative to a naive direct counting code and \textsc{encore}. The input data used was created by randomly placing 104 particles on a grid of resolution $N_{\rm g} = 128^3$. To create the particles on the grid we assign cells within the grid a weight equal to unity. We compare non-normalized coefficients for both the full 3PCF and the full 4PCF to show that the measured coefficients agree. For the full 3/4 PCF we achieve a precision $\sim 10^{-6}$ without normalization as seen in the bottom of each plot. In both plots we show the residual between \textsc{sarabande} and \textsc{encore} in red and the residual between the direct counting code and \textsc{encore} in blue. The precision of agreement between \textsc{sarabande} and the direct counting code is $\sim 10^{-14}$. We use 10 linearly-spaced radial bins for all of the measurements made in this figure. We use the same mapping of bins to a 1D index as discussed in the bottom row of Figure \ref{fig:normalization_projected}.}
\label{fig:full_compare}
\end{figure}

Without normalization \textsc{sarabande} agrees with the particle-based direct counting code to a precision of $\sim 10^{-13}$. When comparing \textsc{sarabande} to \textsc{encore} without normalization we found the full 3/4 PCFs agreed with a precision of $\sim 10^{-6}$ as seen in Figure \ref{fig:full_compare}. After our normalization scheme, the agreement precision of \textsc{sarabande} with particle-based codes increases with resolution. For a grid resolution of $N_{\rm g} = 256^2$ \textsc{sarabande} agrees with the particle-based direct counting code to a precision of $\sim 10^{-3}$ as seen in Figure \ref{fig:normalization_projected}. As the resolution of the grid is increased the precision increases as one would expect because the width of each cell decreases allowing for a better approximation of the analytical bin volume/area (equation \ref{eqn:radial_binning} and equation \ref{eq: proj_b_area}). The same normalization behavior applies to both the full and projected 3/4 PCF in \textsc{sarabande}. The particle-based direct counting code used for testing is publicly available on the \github $\;$ described in Section \ref{sec:code_use}.

% #########################################################
% #########################################################

%                     Code Structure

% #########################################################
% #########################################################
% \vspace{-0.1in}
\section{Code structure and Usage}
\label{sec:code_struc}
\begin{figure*}
\centering
\includegraphics[width=\textwidth]{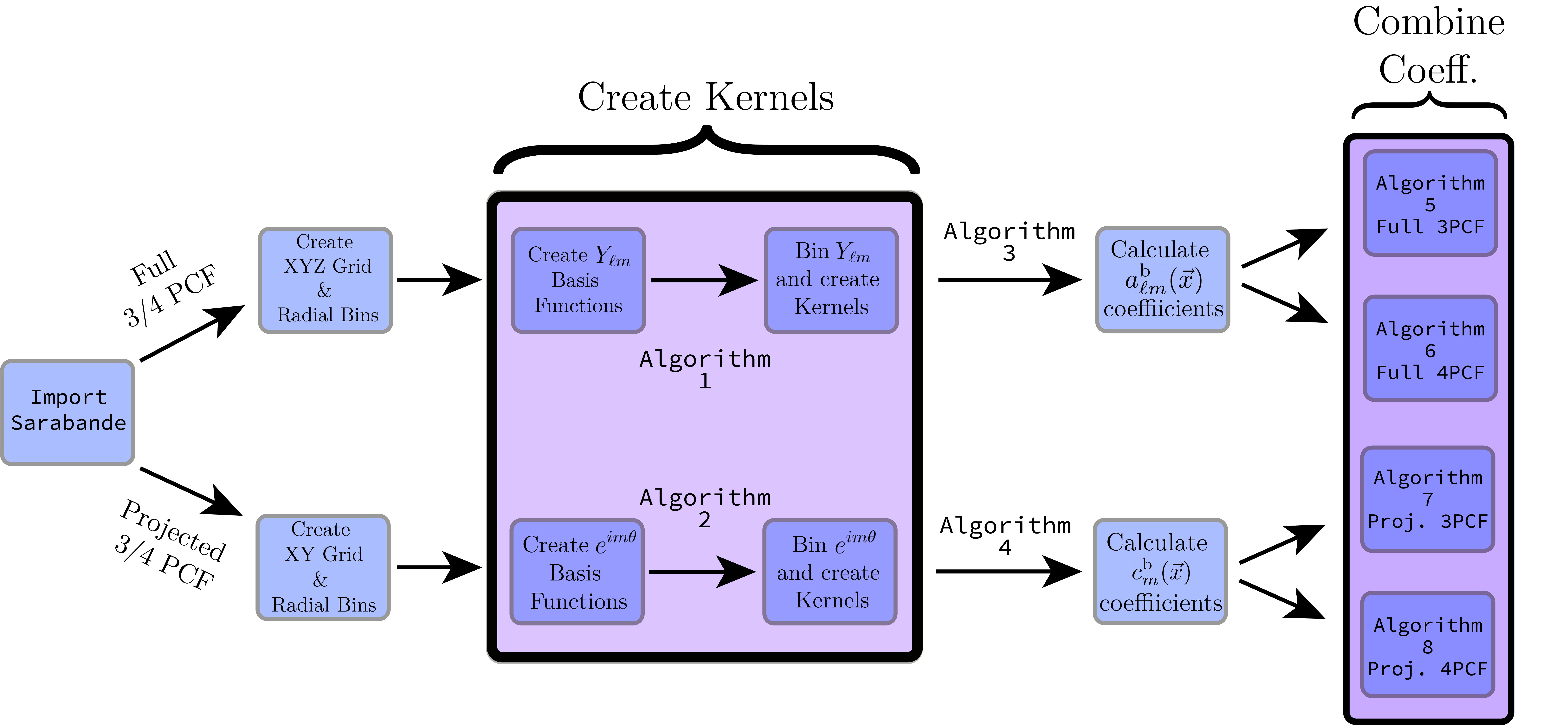}
\caption{Here we show how \textsc{sarabande} proceeds through the calculation of the 3/4 PCFs. The flowchart reads left to right starting with importing the \textsc{sarabande} package. The code splits into two pathways depending on if the user chooses to calculate the projected or full 3/4 PCF. The two pathways then diverge again when deciding to compute the 3 or the 4 PCF (Algorithms \ref{alg:3pcf}-\ref{alg:proj_4pcf}). The code follows a general structure such that we first create kernels out of radially-binned basis functions ($Y_{\ell m}$ or $e^{im\phi}$) which we then convolve with the density field to calculate our convolution coefficients ($a_{\ell m}^{\mathrm{b}} (\vec{x})$ or $c_m^{\mathrm{b}}( \vec{x})$). These convolution coefficients are then combined in the final step of each pathway to give the desired full or projected 3/4 PCF radial coefficients.}
\label{fig:workflow}
\end{figure*}

\begin{figure}
\centering
\includegraphics[width=\columnwidth]{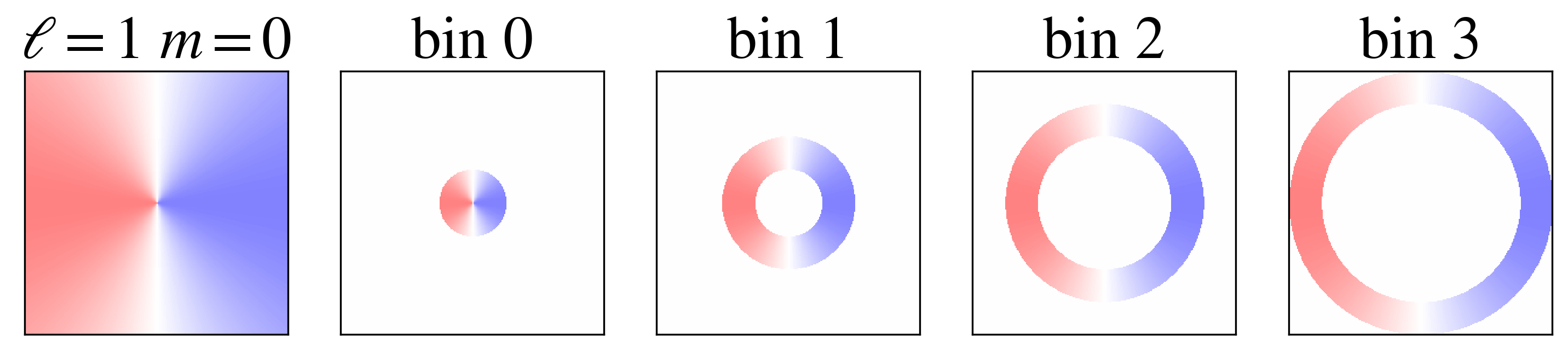}
\caption{Visual example of the binning of spherical harmonics to create kernels for the full 3/4 PCFs. In this figure we have a slice of $Y_{10}$ being binned into 4 separate radial bins numbered 0-3. We convolve these kernels with the original data to create the convolution coefficients.}
\label{fig:binning_}
\end{figure}

\textsc{sarabande} offers two pathways to calculate correlation functions: one pathway is used to calculate full correlation functions, while the other is used to calculate projected correlation functions. Both pathways resemble each other and are outlined in Figure \ref{fig:workflow}. All computations start with creating a grid and radial bins for the input data (2D for projected and 3D for full). From there, both pathways generate kernels based on their basis functions and radial bins as seen in Figure \ref{fig:binning_}. Once the basis kernels are created, then \textsc{sarabande} convolves them with the data to yield the convolution coefficients ($a_{\ell m}^{\mathrm{b}} (\vec{x}$) for 3D and $c_m^{\mathrm{b}}( \vec{x})$ for 2D) necessary for computing the desired correlation function coefficients $\hat{\zeta}$. The final step is to assemble the convolution coefficients into products and sum those products over all cells. We next describe each pathway in more detail in Section \ref{sec:full_pathway} and Section \ref{sec:proj_pathway}; we then give a guide to code use in Section \ref{sec:code_use} and suggest how to explore the code to gain familiarity in Section \ref{sec:explore_code} using an online google colab notebook we have created.
\subsection{Full NPCF Pathway}
\label{sec:full_pathway}
The full 3/4 PCF calculation differs from the projected mostly in the handling of the spherical harmonic kernel calculation. 3D gridded data has a much larger memory footprint than 2D gridded data; thus to hold the computation of $\zeta$ in memory we must offload intermediate calculations to disk and delete them when they are no longer needed. The full pathway is divided into several parts. The pathway starts with the generation of spherical harmonic kernels by radially binning the spherical harmonics as described in Algorithm \ref{alg:part1} and saving their Fourier Transforms. The second part of the pathway convolves a given dataset with the kernels to calculate the convolution coefficients $a_{\ell m}^{\mathrm{b}}(\vec{x})$ as outlined in Algorithm \ref{alg:convolve almb coeffs}. These convolution coefficients are then assembled to yield the full 3/4 PCF as desired. The only difference between calculating the full 3PCF and the full 4PCF is the final algorithm chosen for assembling convolution coefficients. For the full 3PCF we use algorithm \ref{alg:3pcf} which is adapted from \cite{cats_portillo2018} and for the full 4PCF we use algorithm \ref{alg:4pcf} adapted from \cite{philcox2021encore}.

During the entire computation of the full 3PCF we only keep at most three arrays, each of size $\Ngm^3$ in memory at a time, where $\Ngm$ is the number of grid cells on each side of the data grid. This approach requires saving intermediate calculations to disk one array at a time. Only when we need an array for a subsequent calculation is it read back in from disk. When computing the convolution coefficients via Algorithm \ref{alg:convolve almb coeffs}, the three arrays needed for the calculation are the Fourier Transform of the data, the Fourier Transform for a single kernel, and an array to hold the product. Then for the final combination of convolution coefficients in Algorithm \ref{alg:3pcf} the three needed arrays are the density field and two different convolution coefficients. 

For the full 4PCF we only keep up to four arrays of size $\Ngm^3$ in memory at a time. The number of arrays in memory is the same as for the 3PCF except for the final step of combining convolution coefficients. The four arrays in memory during this last step, given by Algorithm \ref{alg:4pcf}, are the density field and three different convolution coefficients. Overall, this limited-memory footprint approach allows \textsc{sarabande} to measure the 3/4 PCFs of high-resolution gridded datasets and simulations.
\subsection{Projected NPCF Pathway}
\label{sec:proj_pathway}

For the projected 3/4 PCF calculation we can afford to keep intermediate calculations in memory due to the much smaller memory footprint of 2D data. This difference inherently means the projected calculations will be much faster than the full 3/4 PCF calculations due to no time being spent on File I/O operations. The projected pathway of \textsc{sarabande} follows the same structure as the full pathway just in 2D instead of 3D. 

The projected pathway also starts with the generation of kernels by radially binning basis functions as described in Algorithm \ref{alg:Create Projected Kernels}. These kernels use a different set of basis functions; we call them the Fourier basis. Specifically, they are complex exponentials, which are essentially a projection of the spherical harmonics onto a 2D plane.\footnote{We recall that $Y_{\ell m} \propto P_{\ell}^m(\cos \theta)\exp[i m \phi]$, with $P_{\ell}^m$ an associated Legendre polynomial. Setting $\cos \theta = 0$ as is the case in the $xy$-plane renders this purely dependent on $\exp[i m\phi]$.} Sections \ref{sec:proj 3pcf} and \ref{sec:proj_4pcf} describe the use of this basis in more detail. Once the kernels have been created the projected pathway convolves them with the density field to calculate the convolution coefficients $c_{m }^{\mathrm{b}}(\vec{x})$ as outlined in Algorithm \ref{alg:Create C_ms}. With these 2D convolution coefficients, we assemble them into the projected 3/4 PCF as desired. We use Algorithm \ref{alg:proj_3pcf} for the projected 3PCF (developed in Section 2.1 of \citealt{saydjari2021}) and Algorithm \ref{alg:proj_4pcf} for the projected 4PCF (which is entirely new). 

% #########################################################
% #########################################################

%                    Guide to Code Use

% #########################################################
% #########################################################
\vspace{-0.1in}
\subsection{Guide to Code Use}
\label{sec:code_use}

\textsc{sarabande} was designed to be simple to use while also allowing the user to execute each step as desired. There are various functions listed in the documentation on \texttt{github}: \github. \footnote{\github: https://github.com/James11222/sarabande} \textsc{sarabande} makes use of object-oriented programming in \texttt{python} to create a \texttt{measure} object. This object can be passed into the zeta calculation function to calculate the desired correlation function as follows: 
% \begin{minted}[frame=lines]{python}
%     import sarabande

%     NPCF_obj = sarabande.measure(**kwargs)
%     sarabande.calc_zeta(NPCF_obj)
%     zeta = NPCF_obj.zeta
% \end{minted}
% \rule{\columnwidth}{0.4pt}
\begin{lstlisting}[language=Python]
    import sarabande

    NPCF_obj = sarabande.measure(**kwargs)
    sarabande.calc_zeta(NPCF_obj)
    zeta = NPCF_obj.zeta
\end{lstlisting}
% \rule{\columnwidth}{0.4pt}

where \texttt{**kwargs} are the possible arguments in the constructor function. There are many arguments and comprehensive definitions for usage that can be found in the documentation at the \texttt{github} repository linked above. The two most important parameters are \texttt{projected [boolean]} and \texttt{NPCF [integer]} as they govern which algorithms will be used to calculate $\zeta$. The variable type is denoted by square brackets. We provide an optional \texttt{normalized [boolean]} flag argument to activate the normalization scheme discussed in Section \ref{sec:normalization}. This calculation is then stored by adding the \texttt{.zeta} attribute to the \texttt{measure} object.

\subsection{Explore the Code}
\label{sec:explore_code}

We have also provided tutorial \textsc{colab} notebooks to show the process of calculating the 3/4 PCFs (projected and full). These notebooks walk through each step of the code interactively without requiring one to examine the source code of \textsc{sarabande}. The notebooks are located here: \href{https://drive.google.com/drive/folders/1oEum7DThj9kkAbyacx0eX-x9qY-8cZv0?usp=sharing}{\textsc{sarabande drive}}. \footnote{https://drive.google.com/drive/folders/1oEum7DThj9kkAbyacx0eX-x9qY-8cZv0?usp=sharing}

% #########################################################
% #########################################################

%                        Results

% #########################################################
% #########################################################

\section{Performance and Validation}
\label{sec:results}

\begin{figure}
\centering
\includegraphics[width=\columnwidth]{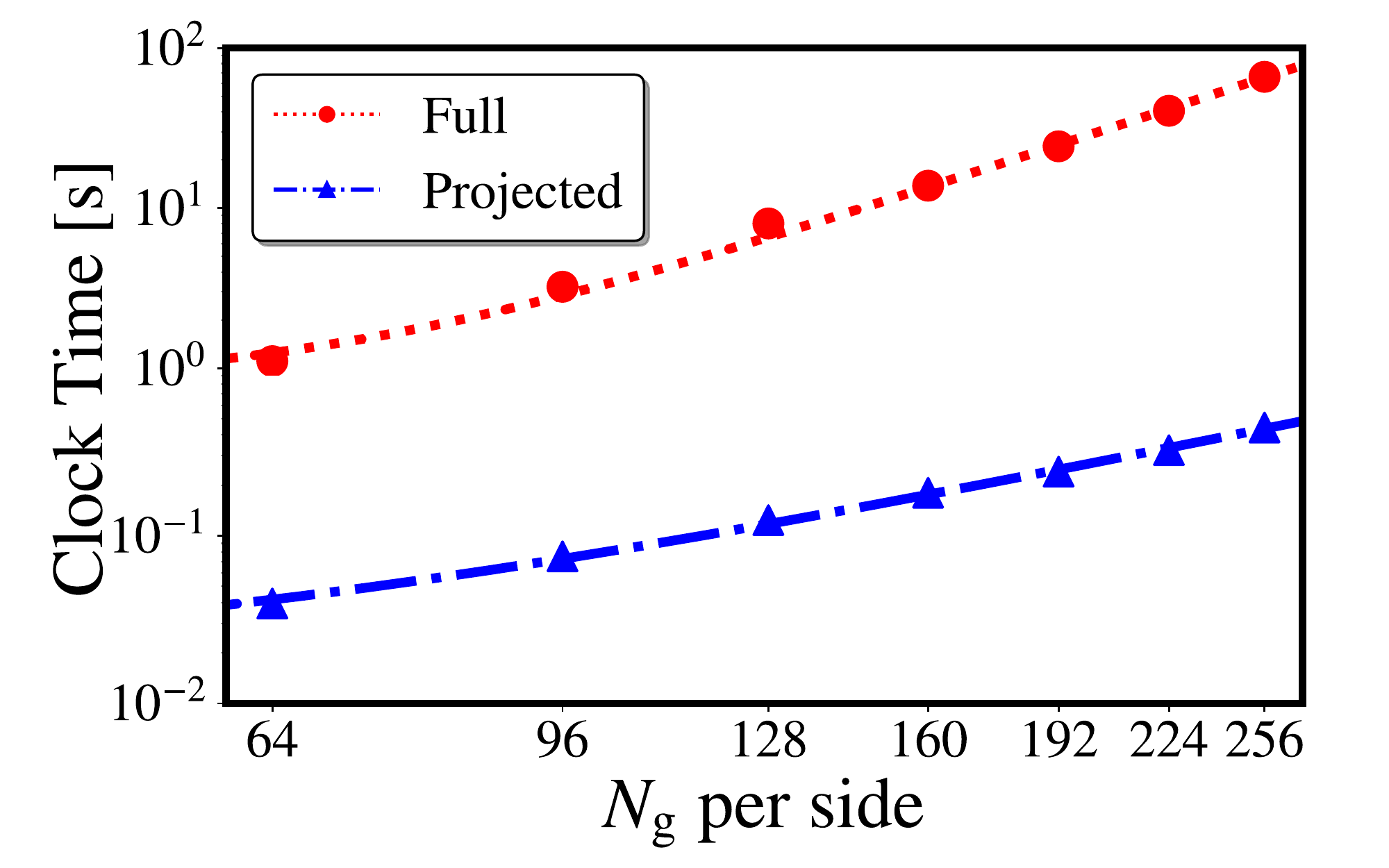}
\caption{Here we show the scaling of the creation of the convolution coefficients in \textsc{sarabande} (Algorithms \ref{alg:part1}-\ref{alg:Create C_ms}) as outlined in Figure \ref{fig:workflow}. This plot shows the scaling in clock time vs. $\Ngm$ per side of a given data set. $\Ngm$ is defined as the total number of grid cells used for the FFT, so the full 3/4 PCFs are using $(\Ngm \; \mathrm{per \; side})^3$ cells and the projected 3/4 PCFs are using $(\Ngm \; \mathrm{per \; side})^2$ cells. Both fitted lines scale as $\Ngm \log(\Ngm)$ which agrees with the predicted scaling of our algorithms for the full and projected 3/4 PCFs. The full 3/4 PCFs are calculated with 4 radial bins and an $\ell_{\textrm{max}} = 1$ and the projected 3/4 PCFs are calculated with 10 radial bins and $m_{\textrm{max}} = 5$ in this example. We omit the computational time spent combining the convolution coefficients (Algorithms \ref{alg:3pcf}-\ref{alg:proj_4pcf}) because these algorithms are strongly dependent on the number of File I/O operations instead of the resolution of the input data. The amount of File I/O operations needed for combining the convolution coefficients is determined by the choice of $N_{\rm bins}$ and $\ell_{\rm max}$. The storage of convolution coefficients on disk is necessary to keep the calculations in memory as discussed in Section \ref{sec:full_pathway}.}
\label{fig:scaling}
\end{figure}

The attractive features of \textsc{sarabande} are that it is written in \texttt{python}, and that it formally scales with resolution as $\mathcal{O}(N_{\rm g} \log N_{\rm g})$, as shown in Figure \ref{fig:scaling}. This scaling is made possible by the use of the FFT algorithm originally discovered by \cite{cooley1965algorithm} which has a complexity scaling of $\mathcal{O}(N_{\rm g} \log N_{\rm g})$. The history of the FFT algorithm is discussed in \cite{Gauss_FFT}. The scaling of \textsc{sarabande} enables computing higher-order correlation functions of high-resolution gridded datasets and simulations. Using a particle-based code such as \textsc{encore} \citep{philcox2021encore} the formal scaling would be $\mathcal{O}(N_\mathrm{p}^2)$, with $N_\mathrm{p}$ the number of particles in the dataset.\footnote{In actual practice for test datasets, \textsc{encore} scales linearly in number of particles for the full 4PCF, because it is not the $a_{\ell m}$ computation that actually dominates the runtime, but assembling the products of coefficients. \textsc{encore}, and the original particle-based 3PCF algorithm \cite{SE_3pt_alg} on which it is based, does scale as $N_{\mathrm{p}}^2$ for 3PCF. \textsc{encore} does not include projected statistics.}   Previously the higher-order statistics provided by the N-point correlation function were primarily applied in a cosmological setting in the context of galaxy redshift surveys \citep{BOSS, SDSS}. There has been some work in the application of the 3PCF and Bispectrum on Magnetohydrodynamic (MHD) simulations of the ISM \citep{cats_portillo2018, blakesly_2022}. \textsc{sarabande} allows us to measure higher-order statistics of any gridded data set, opening the door for N-point statistical analysis in other subfields of astronomy than just cosmology. As an example of this, we show sample measurements made on a simulation of the ISM produced by the CATS collaboration \citep{cats_bialy2020, cats_bkhart2009, cats_cho_lazarian_2003, cats_portillo2018,  cats_burkhart_2020}. 

\begin{figure*}
\centering
\includegraphics[width=\textwidth]{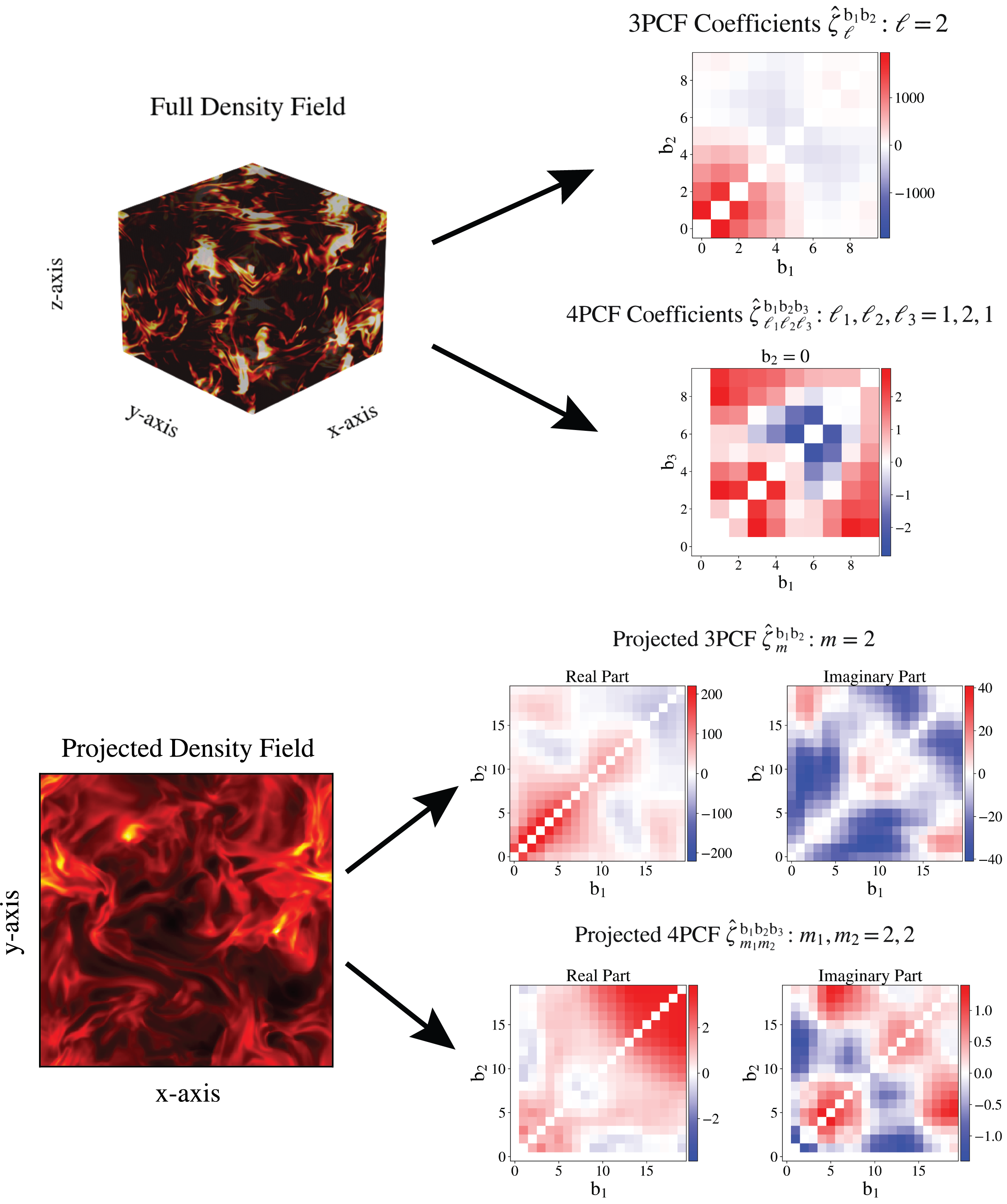}
\caption{
Here we show an example application of \textsc{sarabande}. The top half of the figure shows the results of the full 3/4 PCF algorithm measurements on the 3D density field shown on the left. The bottom half of the figure shows the results of the projected 3/4 PCF algorithm measurement on the projected density field displayed on the left. The full 3/4 PCF coefficients in the top half are purely real while the projected coefficients in the bottom half are complex. The projected 4PCF coefficients shown correspond to $\rm b_3 = 0$. For measurements we subtract off the mean of the input density field so that positive and negative coefficients correspond to excess and deficit relative to random respectively. We used 10 radial bins for the full 3/4 PCFs and 20 radial bins for the projected 3/4 PCFs. For all cases we ignore bin overlap so that the diagonals, where $\mathrm{b}_1 = \mathrm{b}_2$ or $\mathrm{b}_1 = \mathrm{b}_2 = \mathrm{b}_3$, have been set to zero.}
\label{fig: Full_path}
\end{figure*}

\textsc{sarabande} can take either 2D data slices or 3D data cubes as input to measure either projected or full 3/4 PCFs. Figure \ref{fig: Full_path} serves as a typical example output one can expect when using \textsc{sarabande} in practice. The top half of the figure shows example coefficients of a full 3/4 PCF and the bottom half shows the process of measuring a projected 3/4 PCF. The density field used in Figure \ref{fig: Full_path} is from a simulation of the ISM produced using a third-order-accurate hybrid Essentially Non-Oscillatory (ENO) scheme developed in \cite{cats_cho_lazarian_2003} to solve the ideal magnetohydrodynamic equations in a periodic box with driven turbulence. The simulation chosen for Figure \ref{fig: Full_path} is characterized by an average sonic Mach number of $\mathcal{M}_s$$\sim$$1.2$ and an average Alfvénic Mach number of $\mathcal{M}_A$$\sim$$2.0$.

The sonic Mach number is defined as $\mathcal{M}_s \equiv |\vec{v}|/c_s$ and the Alfvénic Mach number to be $\mathcal{M}_A \equiv |\vec{v}|/ \left<v_A\right>$. $\vec{v}$ is the velocity field, $c_s$ the isothermal sound speed, $v_A$ the Alfvén speed, and angle brackets denote a spatial average over the entire simulation box. The full simulation chosen has a resolution of $256^3$ grid cells but for the projected 3/4 PCF we only take a single slice of this density cube such that there are a total of $256^2$ grid cells in the density field. This simulation is one of many that has been described and used in previous work by the CATS collaboration \citep{cats_bialy2020, cats_bkhart2009, cats_cho_lazarian_2003, cats_portillo2018, cats_burkhart_2020} The \href{www.mhdturbulence.com}{CATS Database} webpage gives more details on the simulations. \footnote{CATS Database: www.mhdturbulence.com}

The bottom half of Figure \ref{fig: Full_path} shows example results for measuring the projected 3/4 PCFs using \textsc{sarabande}. The input for the bottom half of this figure is simply a slice taken from the full density cube used in the top half of the figure. For all measurements made we subtract out the mean from the data so that all 3/4 PCF coefficients represent excesses and deficits relative to random, depicted by red and blue in the coefficients respectively. A noticeable difference in the outputs between the full and projected pathways is that the projected 3/4 PCF coefficients are complex-valued while the full 3/4 PCF coefficients are entirely real-valued. Since the projected 3/4 PCFs are far less expensive to measure than their full counterparts we can probe significantly higher resolutions in a fraction of the time. It is apparent that there is structure in the 3/4 PCF coefficients both in the full and projected versions. This structure suggests further study which we are pursuing in a subsequent paper (Sunseri \textit{et al.} in prep).

\begin{figure*}
\centering
\includegraphics[width=2\columnwidth]{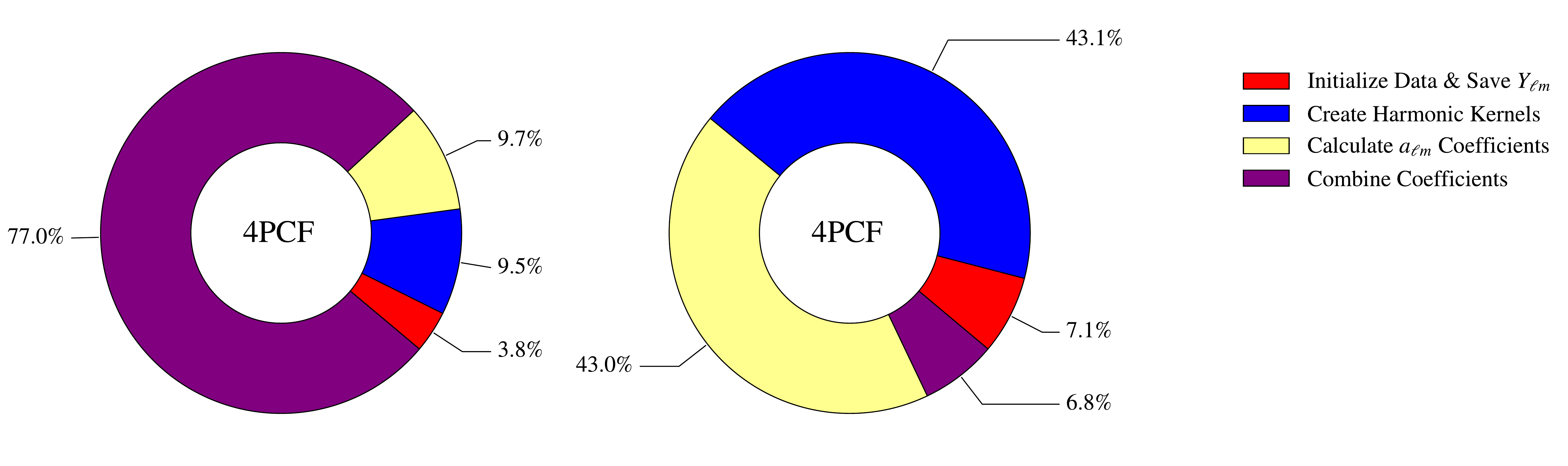}
\caption{Here we show the impact of parallelizing Algorithm \ref{alg:4pcf} for the full 4PCF. The left pie chart shows the distribution of time spent computing the full 4PCF without any parallelization. The right pie chart shows the same calculation where we include parallelization of Algorithm \ref{alg:4pcf}. A substantial amount of time is spent on the combination of convolution coefficients due to the inefficiency of having a single processor reading in convolution coefficient files from disk within 8 nested \texttt{for} loops. By unraveling the loops with a lookup table we were able to distribute the loops amongst multiple processors. With parallelization we see that the FFTs used to create the convolution coefficients are the rate-limiting factor. The timings displayed above are made using 5 radial bins and an $\ell_{\textrm{max}} = 2$.}
\label{fig:compare_parallel}
\end{figure*}

\subsection{Performance}
\textsc{sarabande} is a valuable tool due to the improved computational speed gained by using FFTs as opposed to direct counting methods for measuring 3/4 PCFs. In Figure \ref{fig:scaling} we show how the creation of convolution coefficients scale in clock time as a function of resolution of the input data (number of cells $\Ngm$). In Figure \ref{fig:scaling} we fit the timing data to scalings $\propto \Ngm \log \Ngm$; this is the expected scaling that occurs for both the projected and the full 3/4 PCFs. We do not include the coefficient combination algorithms (Algorithms \ref{alg:3pcf}-\ref{alg:proj_4pcf}) in this plot because they are primarily dependent on the number of radial bins chosen and the maximum order of angular multipole rather than the grid resolution of the density field.

As seen in Figure \ref{fig:compare_parallel}, it is clear that without multiprocessing the full 4PCF is dominated by the combination of convolution coefficients rather than their creation. Because of this, we parallelize this task of combining coefficients in the 8 nested \texttt{for} loops of Algorithm \ref{alg:4pcf} by unraveling the loops via multi-process mapping using a lookup table approach. This approach allows for a drastic improvement in overall time spent computing the 4PCF depending on how many processors are available. As an example, for a 2.9 GHz Quad-Core Intel Core i7 Macbook Pro we saw a roughly factor of 4 speed-up for Algorithm \ref{alg:4pcf} and expect performance to increase with the number of processors available for the full 4PCF. 

\begin{figure*}
\includegraphics[width=\textwidth]{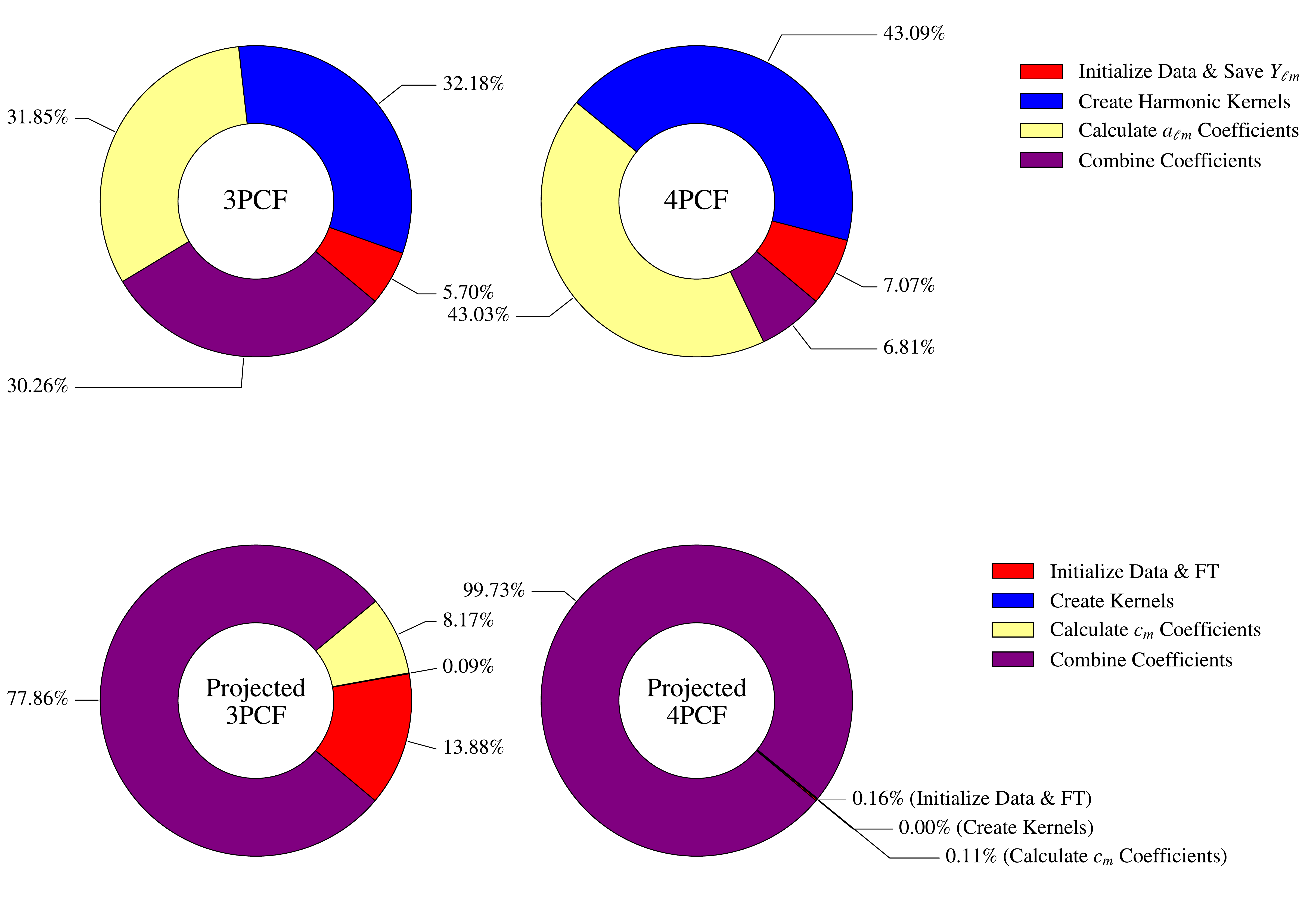}
\centering
\caption{Here we show an example of the fraction of time spent for each of the different computations in the code with a given set of code parameters. For the full 3/4 PCFs, the computational time is dominated by the computation of the convolution coefficients (Algorithms \ref{alg:part1} and \ref{alg:convolve almb coeffs}) when our choice of $\ell_{\rm max}$ and $N_{\rm bins}$ are small. As $\ell_{\rm max}$ and $N_{\rm bins}$ increase, the number of File I/O operations increases causing the total computational time of the full 3/4 PCFs to become dominated by combining the coefficients (Algorithms \ref{alg:3pcf} and \ref{alg:4pcf}). The computation of the convolution coefficients in \textsc{sarabande} scales in complexity with the number of grid cells used in the input data as $\mathcal{O}(N_{\rm g} \log N_{\rm g})$ as shown in Figure \ref{fig:scaling}. This scaling is important because this part of the total computation is dominant for high resolution data measurements. The total computation time of the projected 3/4 PCFs is dominated by the combination of convolution coefficients (Algorithms \ref{alg:proj_3pcf} and \ref{alg:proj_4pcf}). It is expected that the computation of the convolution coefficients (Algorithms \ref{alg:Create Projected Kernels} and \ref{alg:Create C_ms}) does not dominate the total computation time for the projected 3/4 PCFs because the number of grid cells $N_{\rm g}$ is significantly smaller in 2D compared to 3D. This reduction in the number of grid cells of the input data allows us to avoid File I/O operations entirely. The combination of convolution coefficients is more dominant in the projected 4PCF than the projected 3PCF due to the additional nested for loops. The creation of kernels is nearly negligible for the projected pathway, while it is more significant in the full pathway. For these timings we used we used $N_{\rm bins} = 5$, $\ell_{\textrm{max}} = 2$, and a grid resolution of $N_{\rm g} = 128^3$ for the full 3/4 PCFs. For the timing of the projected computation we used $N_{\rm bins} = 5$, $m_{\textrm{max}} = 5$, and a grid resolution of $N_{\rm g} = 128^2$. }
\label{fig:pie chart}
\end{figure*}

In Figure \ref{fig:pie chart} we provide a breakdown of the computational time spent for each of the four types of measurements for a given set of parameters. It is clear that the rate limiting step for the projected 3/4 PCFs is not the calculation of the convolution coefficients, but instead the combination of the convolution coefficients to create the correlation function coefficients. The convolution coefficients are calculated quickly because intermediate results are held in memory instead of on disk, hence the slowest part of the pathway for the projected 3/4 PCFs is the nested \texttt{for} loops in Algorithms \ref{alg:proj_3pcf} and \ref{alg:proj_4pcf}.

As discussed in Sections \ref{sec:full_pathway} and \ref{sec:proj_pathway}, \textsc{sarabande} offloads intermediate calculations to disk only in the full pathway due to the large memory footprint of 3D arrays. The projected pathway outlined in Section \ref{sec:proj_pathway} only requires 2D arrays, which demand a far smaller amount of memory, allowing us to keep the entire computation in memory without offloading calculations to disk. The full 3PCF only requires three $N_{\rm g}^3$ arrays in memory at once: the kernel FFT, the data FFT, and their product. If we take $N_{\rm g}$ to be 256 then we would expect each array to occupy $256^3 \times 2 \times 8$ bytes = 268 MB if we use complex double-precision arrays. Three of these arrays in memory at any given time leads to a maximum memory usage of $3 \times 268$ MB = 0.81 GB total. Similarly, for $N_{\rm g} = 512$ we expect a required total memory of 6.44 GB, and $N_{\rm g} = 1024$ we would need 51.54 GB of memory. For the full 4PCF the only difference is that we have at most four $N_{\rm g}^3$ arrays in memory at once instead of three: three convolution coefficients and the density field. In this case then if we have  $N_{\rm g} = 256$ we would expect $4 \times 268$ MB = 1.07 GB. For $N_{\rm g} = 512$ we expect a total memory usage of 8.59 GB and for $N_{\rm g} = 1024$ we expect a total memory usage of 68.72 GB.

The memory footprint is much smaller for the projected 3/4 PCF. If we use an $N_{\rm g} = 256$ in this case then we would expect the $N_{\rm g}^2$ complex double-precision arrays to require  $256^2 \times 2 \times 8$ bytes = 1.04 MB per array. Even for a resolution of $N_{\rm g} = 1024$, we expect each array to use 16.78 MB of memory. This small memory footprint allows us to store all convolution coefficients in memory at any given time without offloading to disk. The number of these $N_{\rm g}^2$ arrays in memory is determined by the number of bins used and the maximum value of $m$. The total memory required is equal to $(\textrm{Memory per Array}) \times (N_{\rm bins} \times m_{\rm max} + 1)$. For a resolution of $N_{\rm g} = 1024$ we expect to use 1.69 GB of memory total to store all the necessary arrays for $N_{\rm bins} = 20$ and $m_{\rm max} = 5$. 

% When comparing the computational time required t

\subsection{Validation}

To validate our code we compare \textsc{sarabande} to other codes that measure the 3/4 PCFs. For the full 3/4 PCFs we compare our code to both the current standard particle-based code \textsc{encore} and a naive direct counting code which simply counts triplets or quadruplets respectively. These codes should all be equivalent for the same input data since they are measuring the same correlation functions. Since \textsc{encore} does not compute projected 3/4 PCFs, we can only compare our code to the naive direct counting code in this case. 

To generate an equivalent input for both the particle-based codes and our grid based code we start with an empty grid and randomly populate cells of this grid with a weight of unity. We then treat the center of each of these populated cells as the coordinate of the "particles" within the box. The box is defined to have side length equal to the resolution of the grid (\textit{i.e} a grid of resolution $128^3$ will correspond to a box of side length $128$). \textsc{sarabande} takes the original grid as its input while \textsc{encore} and the direct counting code both take in a list of coordinates for the particles. These inputs are inherently not precisely the same so we do expect minor discrepancies that increase in magnitude with lower resolution of the mesh. As the resolution increases our cells are better approximated as point particles. We give examples of how the codes compare in Figure \ref{fig:normalization_projected} and Figure \ref{fig:full_compare}. It is evident from these figures that the codes are measuring the same correlation coefficients except with minor disagreement. It is expected that with a higher resolution of the grid these discrepancies will approach machine error precision. 

\section{Future Applications}
\label{sec:Future Apps}

There have been previous efforts to use both the full and projected 3PCF to better understand grid-based simulations of the turbulent ISM
\citep{cats_portillo2018, saydjari2021}. The present work serves as a release of open software available for making measurements of both full and projected 3/4 PCFs. In a future study (Sunseri \textit{et al.} 2022 in prep.) we plan to report the first application of the full and projected 4PCF on the ISM and other possible applications using \textsc{sarabande}. 

In addition to applications at galactic scales, \textsc{sarabande} can also be applied to datasets on cosmological scales. As discussed elsewhere, there have been several previous applications of full 3PCF and 4PCF to galaxies; here we outline future possibilities. \textsc{sarabande} can be applied to galaxy redshift surveys by obtaining either spectroscopic redshifts~\citep{BOSScollaboration2017,eBOSScollaboration2021,DesiCollaboration2016} or photometric redshifts~\citep{LSST:2012}. In addition to surveys with individually-resolved galaxies, \textsc{sarabande} can also be applied to surveys using unresolved galaxies or the diffuse intergalactic medium~\citep{Nan2011fast, vanHaarlem2013lofar,Dore2014spherex,Shaw2014chime,Battye2016bingo} by measuring the intensity of a chosen emission line. These surveys usually cover a large sky fraction but with relatively low resolution compared to surveys that resolve individual galaxies. Within each redshift shell, the measured intensity is a projected quantity; therefore, \textsc{sarabande} with its projected 3/4 PCF implementation is useful for such surveys.

Finally, in many cosmology settings, one computes the desired clustering statistic on many mock catalogs to estimate covariance. Often these mocks are grid-based and here using \textsc{sarabande} to evaluate the covariance would be lossless, and the speed enabling. This would aid, for instance, a DESI 3PCF or 4PCF search for BAO (already detected in the 3PCF and bispectrum of DESI's predecessor Sloan Digital Sky Survey Baryon Oscillation Spectroscopic Survey (BOSS), \citealt{SE_BOSS_3}, \citealt{SE_FULL_3PCF_BAO}, \citealt{pearson}). Especially given the development of fast methods for other parts of the BAO fitting pipeline, \textit{e.g.} \citet{hansen}, speeding up the covariance calculation, which is often rate-limiting, is worthwhile. Given that higher-order statistics such as 3PCF also have been shown to carry extra information on the neutrino mass (\citealt{chang_1}, \citealt{chang_2}, \citealt{fk_nu}, \citealt{aviles}) and on modified gravity \citep{mg_wp}, two other major goals of DESI, new fast algorithms such as \textsc{sarabande} are desirable.

% #########################################################
% #########################################################

%                        Conclusion

% #########################################################
% #########################################################

\section{Conclusion}
\label{sec:conclusion}
\textsc{sarabande} is a new \texttt{python} package for measuring 3 and 4 Point Correlation Functions on gridded data using Fast Fourier Transforms. The use of FFTs gives \textsc{sarabande} a complexity scaling of $\mathcal{O} (\Ngm \log \Ngm)$ which offers the fastest method for measuring the 3/4 PCF to date. The package allows for measuring the standard full 3/4 PCFs on a 3D data cube while also giving the option to measure their projected counterparts on a 2D data slice. Since this package operates on gridded data, it allows users to measure the 3/4 PCF in new contexts yet to be explored. The standard algorithms use a particle-based approach by directly counting triplets and quadruplets, which is slower than our FFT based approach which computes the contributions of triplets or quadruplets simultaneously at every point instead. This older direct-counting method was developed in the context of cosmological surveys where galaxies are treated as points in a survey volume. Now with \textsc{sarabande} we demonstrate how one can measure the 3/4 PCF and their projected counterparts on datasets outside of cosmology. We give an example measurement on a simulation of the the turbulent ISM provided by the CATS collaboration \citep{cats_bialy2020, cats_bkhart2009, cats_cho_lazarian_2003, cats_portillo2018}. 

In the future we plan to explore how the 3 and 4 Point Correlation Functions can be applied to MHD simulations to better understand turbulence, shocks, and other physical processes. We also plan to continue optimizing the code by implementing more parallelization schemes and potentially harnessing GPUs for further acceleration (\textit{e.g.} \cite{GPUs}). 
% \vspace{-0.1in}
\section*{Acknowledgements}

We acknowledge the University of Florida Research Computing for providing computational resources and support that have contributed to the research results reported in this publication. We would like to thank Oliver Philcox, Andrew Saydjari, and Matt Hansen for insightful discussions pertaining to the development and optimization of \textsc{sarabande}. Support from Isabella Dougas and Ronan Hix was crucial during the development of \textsc{sarabande}. Lastly, we thank the rest of the Slepian Research Group at University of Florida for their feedback and support on the project. Jiamin Hou has received funding from the European Union’s Horizon 2020 research and innovation program under the Marie Sk\l{}odowska-Curie grant agreement No 101025187. This research would not be possible without the funding provided by the National Science Foundation for the Research Experience for Undergraduates at University of Florida.
% \vspace{-0.1in}
\section*{Data Availability}
All data used for this work are publicly available and can be found on the \github.\footnote{\github: https://github.com/James11222/sarabande}
% \clearpage
\bibliographystyle{rasti}
\bibliography{3pcf_ft_code}

%\begin{thebibliography}{99}

%\end{thebibliography}

%%%%%%%%%%%%%%%%%%%%%%%%%%%%%%%%%%%%%%%%%%%%%%%%%%

% #########################################################
% #########################################################

%                       Algorithms

% #########################################################
% #########################################################

% \newpage
% \clearpage
\appendix
\section*{Appendix}
\label{sec:algorithms}

Here we display each algorithm used in \textsc{sarabande}. Algorithms \ref{alg:part1}-\ref{alg:Create C_ms} are the algorithms used to prepare kernels from basis functions and generate convolution coefficients. Algorithms \ref{alg:3pcf}-\ref{alg:proj_4pcf} are the algorithms used to compile convolution coefficients into their corresponding correlation functions. All algorithms are written heuristically for clarity. For details of the \texttt{python} version, visit the \github. \footnote{\github: https://github.com/James11222/sarabande}

% \url{https://colab.research.google.com/drive/1RjuM-V-Wby2Blzr9EaPv-T4TgW22mZPp}

% #########################################################
%               Algorithm 1: YLM Kernels
% #########################################################

\begin{algorithm}
\KwResult{The creation of $\Ngm \times \Ngm \times \Ngm$ arrays indexed by degree $\ell$, order $m$, and bin index $\mathrm{b}$. The arrays that represent the Fourier Transform of the binned spherical harmonic kernels are then saved to disk.}
\vspace{0.1cm}
\textbf{Notes:} We initialize three $\Ngm \times \Ngm \times \Ngm$ arrays $x,y,z$ containing those coordinates centered on zero then use these arrays to compute spherical harmonics from Cartesian representations.
\vspace{0.1cm}
\\
\SetAlgoLined
 \For{$\ell$ = 0 to $\ell_{\rm max}$}{
  \For{m = 0 to $\ell$}{
  $Y_{\ell m} \leftarrow Y_{\ell m}(x,y,z)$\;
   \For {$\mathrm{b}$ = 1 to $N_{\rm bins}$}{
   $\Phi^\mathrm{b} \leftarrow r_{\rm min}(\mathrm{b}) < \sqrt{x^2 + y^2 + z^2} \leq r_{\rm max}(\mathrm{b})$\;
%   $Y_{\ell m}^{\mathrm{b}} \leftarrow Y_{\ell m} \times \Phi^\mathrm{b}$\;
   $\tilde{Y}_{\ell m}^{\mathrm{b}} \leftarrow FFT(Y_{\ell m} \times \Phi^\mathrm{b})$\;
   save $\tilde{Y}_{\ell m}^{\mathrm{b}}$ to disk\;
   }
  }
 }
 \caption{Generating Full 3/4 PCF Kernels}
 \label{alg:part1}
\end{algorithm}

% #########################################################
%               Algorithm 2: Create e^imθ Kernels
% #########################################################

\begin{algorithm}

\caption{Generate Projected 3/4 PCF Kernels}
\KwResult{The creation of $\Ngm\times \Ngm$ arrays that are indexed by the angular index $m$ and the radial bin index $\mathrm{b}$. These arrays represent the Fourier Transform of the binned Fourier basis kernels.}
\vspace{0.1cm}
\textbf{Notes:} we initialize two $\Ngm\times \Ngm$ arrays $x,y$ containing those coordinates centered on zero, which we use in the following identity for our basis: $e^{i m \phi} = \left(\frac{x + iy}{r}\right)^m$
\vspace{0.1cm}
\\
\SetAlgoLined
\For{$m = 0$ to $m_{\mathrm{max}}$}{
    \For{$\mathrm{b} = 1$ to $N_{\mathrm{bins}}$}{
        $\Phi^\mathrm{b} \leftarrow r_{\rm min}(\mathrm{b}) < \sqrt{x^2 + y^2} \leq r_{\rm max}(\mathrm{b})$\;
        
       $K_m^\mathrm{b} \leftarrow FFT(e^{im\phi} \times \Phi^\mathrm{b}$) \;
    }
}
\label{alg:Create Projected Kernels}
\end{algorithm}

% #########################################################
%               Algorithm 3: Create a_lms
% #########################################################

\begin{algorithm}
\SetAlgoLined
\KwResult{Convolve binned spherical harmonic kernels to construct the $a_{\ell m}^{\mathrm{b}}$ coefficients and save them to disk.}

\vspace{0.1cm}
\textbf{Notes:} The $a_{\ell m}^{\mathrm{b}}$ coefficient arrays have the shape $\Ngm\times \Ngm \times \Ngm$ once indexed by $\ell, m, \mathrm{b}$.
\vspace{0.1cm}
\\
 $\delta \leftarrow$ read in density field\;
 $\tilde{\delta} \leftarrow FFT(\delta)$
 
 \For{$\ell$ = 0 to $\ell_{\rm max}$}{
  \For{m = 0 to $\ell$}{
   \For {$\mathrm{b}$ = 1 to $N_{\rm bins}$}{
    read $\tilde{Y}_{\ell m}^{\mathrm{b}}$ from disk\;
    $a_{\ell m}^{\mathrm{b}}\leftarrow FFT^{-1}(\tilde{\delta} \times \tilde{Y}_{\ell m}^{\mathrm{b}})$\;
    save $a_{\ell m}^{\mathrm{b}}$ to disk\;
   }
  }
 }
 \caption{Convolve Full 3/4 PCF Kernels}
 \label{alg:convolve almb coeffs}
\end{algorithm}

% #########################################################
%               Algorithm 4: Create c_ms
% #########################################################

\begin{algorithm}

\caption{Convolve Projected 3/4 PCF Kernels}
\KwResult{Convolve the data with the projected kernels to create $c_m^\mathrm{b}$ coefficients which can be indexed by $m$ (angular index) and $\mathrm{b}$ (radial bin index).}
\vspace{0.1cm}
\textbf{Notes:} $c_m^\mathrm{b}$ coefficient arrays have the shape $\Ngm \times \Ngm$ once indexed by $\mathrm{b}$ and $m$.
\vspace{0.1cm}
\\
\SetAlgoLined
$\delta \leftarrow$ read in density field\;
$\tilde{\delta} \leftarrow FFT(\delta)$\;
\For{$m = 0$ to $m_{\mathrm{max}}$}{
    \For{$\mathrm{b} = 1$ to $N_{\mathrm{bins}}$}{
        $c_m^\mathrm{b} = FFT^{-1} (\tilde{\delta} \times K_m^\mathrm{b})$
    }
}
\label{alg:Create C_ms}
\end{algorithm}

% #########################################################
%               Algorithm 5: Calc 3PCF Coefficients
% #########################################################

\begin{algorithm}

\caption{Combine Full 3PCF Coefficients}
\KwResult{Computation of the \textit{unique} full 3PCF coefficients from the $a_{\ell m}^{\mathrm{b}}$ coefficients computed in algorithm \ref{alg:convolve almb coeffs}. The end result is a 3D array indexed by $\ell, \mathrm{b}_1, \mathrm{b}_2$ where $\ell$ is the order of the angular component of the triangles and $\mathrm{b}_1$ and $\mathrm{b}_2$ denote the radial bin length of side 1 and 2 of the triangles respectively. }
\vspace{0.1cm}
\textbf{Notes:} We only calculate the \textit{unique} coefficients since the structure of the full 3PCF is symmetric under permutations of the indices $\mathrm{b}_1$ and $\mathrm{b}_2$. To recover the full 3PCF one can just permute indices and preserve corresponding values.
\vspace{0.1cm}
\\
\SetAlgoLined
$\delta \leftarrow$ read in density cube\\
initialize $\zeta$ as an $(\ell_{\rm max}+1)\times N_{\rm bin}\times N_{\rm bin}$ array of zeros\;
 \For{$\ell$ = 0 to $\ell_{\rm max}$}{
  \For{$\mathrm{b}_1$ = 1 to $N_{\rm bins}$}{
   \For {m = 0 to $\ell$}{
   load $a_{\ell m}^{\mathrm{b}_1}$ from disk\;
    \For {$\mathrm{b}_2$ = 1 to $\mathrm{b}_1$}{
     load $a_{\ell m}^{\mathrm{b}_2}$ from disk\;
     S $\leftarrow$ sum$\left(\delta \times a_{\ell m}^{\mathrm{b}_1} \times a_{\ell m}^{\mathrm{b}_2 *}\right)$\;
     
     \If {m > 0}{
     $S$ +=  $S^*$\;
     }
     $\zeta$[$\ell$, $\mathrm{b}_1$, $\mathrm{b}_2$] += S\;
    }
   }
  }
 }

\label{alg:3pcf}
\end{algorithm}

% #########################################################
%               Algorithm 6: Calc 4PCF Coefficients
% #########################################################

\begin{algorithm}

\caption{Combine Full 4PCF Coefficients}
\KwResult{
% We calculate the unique 4PCF coefficients with this algorithm. In order to get the full 4PCF one can simply permute the indices of their $\zeta^R_\Lambda$ array and set them equal to the previous permutation. We do this to save computation time in the nested loops.
Computation of the \textit{unique} full 4PCF coefficients from the $a_{\ell m}^\mathrm{b}$ coefficients computed in algorithm \ref{alg:convolve almb coeffs}. The end result is a 6D array indexed by $\ell_1, \ell_2,\ell_3, \mathrm{b}_1, \mathrm{b}_2, \mathrm{b}_3$ where $\ell_i$ is the order of the $i$th angular component of the tetrahedrons and $\mathrm{b}_1, \mathrm{b}_2$ and $\mathrm{b}_3$ denote the radial bin length of each side of the tetrahedrons respectively. }
\vspace{0.1cm}
\textbf{Notes:} We only calculate the \textit{unique} coefficients since the structure of the full 4PCF is symmetric under permutations of the indices ($\ell_1,\ell_2,\ell_3)$ and ($\mathrm{b}_1,\mathrm{b}_2, \mathrm{b}_3$). To recover the full 4PCF one can just permute indices and preserve corresponding values. We also define a small function to reduce clutter with if statements below. 
\\
$$
\textrm{define } S(m)= 
\begin{cases}
    1/2 & \text{if } m = 0\\
    1               & \text{else}
\end{cases}
$$
\\
\vspace{0.1cm}
\SetAlgoLined
$\delta \leftarrow$ read in density cube\\
 \For{$\ell_1$ = 0 \rm{to} $\ell_{\rm max}$}{
  \For{$\ell_2$ = 0 \rm{to} $\ell_{\rm max}$}{
   \For{$\ell_3 = |\ell_1 - \ell_2|$ \rm{to} $\rm min(\ell_1 + \ell_2, \ell_{\rm max})$}{
    \If{$\ell_1 + \ell_2 + \ell_3 \mathrm{\; is \; odd}$}{\textbf{continue}}
    \For{$m_1 = -\ell_1$ \rm{to} $\ell_1$}{
     \For{$m_2 = -\ell_2$ \rm{to} $\ell_2$}{
     $m_3 = -(m_1 + m_2)$\\
     \If{$m_3 > \ell_3$ \bf{or} $m_3 < 0$}{\textbf{continue}}
      $\mathcal{W} = \delta(\vec{x}) C^\Lambda_M$\\
      \For{$\mathrm{b}_1 = 1$ \rm{to} $N_{\rm{bins}}$}{
       \eIf{$m_1 > 0$}{
       load $a_{\ell_1 m_1}^{\mathrm{b}_1}$}{load $(-1)^{m_1} a_{\ell_1 m_1}^{\mathrm{b}_1 *}$}
       \For{$\mathrm{b}_2 = \mathrm{b}_1 + 1$ \rm{to} $N_{\rm{bins}}$}{
        \eIf{$m_1 > 0$}{
       load $a_{\ell_2 m_2}^{\mathrm{b}_2}$}{load $(-1)^{m_2} a_{\ell_2 m_2}^{\mathrm{b}_2 *}$}
        \For{$\mathrm{b}_3 = \mathrm{b}_2 + 1$ \rm{to} $N_{\rm{bins}}$}{
        load $a_{\ell_3 m_3}^{\mathrm{b}_3}$\\
        $\zeta[\ell_1, \ell_2,\ell_3, \mathrm{b}_1, \mathrm{b}_2, \mathrm{b}_3] \mathrel{+} = \textrm{sum} ( 2S(m_3)\mathcal{W}\times 
        \textrm{Re}[a_{\ell_1 m_1}^{\mathrm{b}_1} a_{\ell_2 m_2}^{\mathrm{b}_2} a_{\ell_3 m_3}^{\mathrm{b}_3}] )$

        }
       }
      }
     }
    }
   }
  }
 }

\label{alg:4pcf}
\end{algorithm}

% #########################################################
%           Algorithm 7: Calc Projected 3PCF
% #########################################################

\begin{algorithm}

\caption{Combine Projected 3PCF Coefficients}
\KwResult{Compute the \textit{unique} final 3PCF coefficients by combining all the $c_m^\mathrm{b}$ convolution coefficients. The final result is a 3D array which can be indexed by $\mathrm{b}_1, \mathrm{b}_2$ and $m$ where $\mathrm{b}_1$ and $\mathrm{b}_2$ are the radial bin indices for each side of the projected triangles, and $m$ is the index capturing the angular multipole.}
\vspace{0.1cm}
\textbf{Notes:} We only calculate the \textit{unique} coefficients since the structure of the projected 3PCF is symmetric under permutations of the indices $\mathrm{b}_1$ and $\mathrm{b}_2$. To recover the full projected 3PCF one can just permute indices and preserve corresponding values.
\vspace{0.1cm}
\\
\SetAlgoLined

\For{$\mathrm{b}_1 = 1$ to $N_{\mathrm{bins}}$}{
    \For{$\mathrm{b}_2 = 1$ to $\mathrm{b}_1$}{
        \For{$m = 0$ to $m_{\mathrm{max}}$}{
            $\zeta_{\mathrm{proj}}[\mathrm{b}_1, \mathrm{b}_2, m]$ += sum$\left(\delta (\vec{x}) \times c_m^{\mathrm{b}_1} c_m^{\mathrm{b}_2 *} \right)$
        
        }
    }
}

\label{alg:proj_3pcf}
\end{algorithm}

% #########################################################
%           Algorithm 8: Calc Projected 4PCF
% #########################################################

\begin{algorithm}

\caption{Combine Projected 4PCF Coefficients}
\KwResult{Compute the \textit{unique} final 4PCF coefficients by combining all the $c_m^\mathrm{b}$ convolution coefficients. The final result is a 5D array which can be indexed by $\mathrm{b}_1, \mathrm{b}_2, \mathrm{b}_3, m_1$ and $m_2$ where $\mathrm{b}_1,\mathrm{b}_2$ and $\mathrm{b}_3$ are the radial bin indices for each side of the projected tetrahedrons, and $m_1,m_2$ are the indices capturing the angular multipoles of each free angle.}
\vspace{0.1cm}
\textbf{Notes:} We only calculate the \textit{unique} coefficients since the structure of the projected 4PCF is symmetric under permutations of the indices ($\mathrm{b}_1,\mathrm{b}_2, \mathrm{b}_3$) and ($m_1, m_2$). To recover the full projected 4PCF one can just permute indices and preserve the corresponding values. We also do not index by $m_3$ due to the constraint $m_3 = -(m_1 + m_2)$.
\vspace{0.1cm}
\\
\SetAlgoLined

\For{$\mathrm{b}_1 = 1$ to $N_{\mathrm{bins}}$}{
    \For{$\mathrm{b}_2 = 1$ to $\mathrm{b}_1$}{
        \For{$\mathrm{b}_3 = 1$ to $\mathrm{b}_2$}{
            \For{$m_1 = -m_{\mathrm{max}}$ to $m_{\mathrm{max}}$}{
                \If{$m_1 < 0$}{
                   $c_{m_1}^{\mathrm{b}_1} = (-1)^{m_1} c_{-m_1}^{\mathrm{b}_1}$ 
                }
                \For{$m_2 = -m_{\mathrm{max}}$ to $m_{\mathrm{max}}$}{
                    \If{$m_2 < 0$}{
                   $c_{m_2}^{\mathrm{b}_2} = (-1)^{m_2} c_{-m_2}^{\mathrm{b}_2}$ 
                    }
                    $m_3 = -(m_1 + m_2)$
                    
                    \eIf{ $\mathrm{abs}(m_3) \geq m_{\mathrm{max}}$}{
                        continue
                    }{
                        \If{$m_3 < 0$}{
                   $c_{m_3}^{\mathrm{b}_3} = (-1)^{m_3} c_{-m_3}^{\mathrm{b}_3}$ 
                        }
                        $\zeta_{\mathrm{proj}}[\mathrm{b}_1, \mathrm{b}_2,  \mathrm{b}_3,m_1, m_2]$ += sum$\left( \delta(\vec{x}) \times c_{m_1}^{\mathrm{b}_1} c_{m_2}^{\mathrm{b}_2} c_{m_3}^{\mathrm{b}_3} \right)$

                    }

                }
            }
        }
    }
}
\label{alg:proj_4pcf}
\end{algorithm}

%%%%%%%%%%%%%%%%%%%%%%%%%%%%%%%%%%%%%%%%%%%%%%%%%%

%%%%%%%%%%%%%%%%% APPENDICES %%%%%%%%%%%%%%%%%%%%%

%\appendix

%\section{if needed}

%%%%%%%%%%%%%%%%%%%%%%%%%%%%%%%%%%%%%%%%%%%%%%%%%%

% Don't change these lines
\bsp	% typesetting comment
\label{lastpage}
\end{document}